\documentclass[acmsmall,screen,nonacm]{acmart}

\AtBeginDocument{}

\setcopyright{cc}
\setcctype{by}
\acmDOI{10.1145/3839519}
\acmYear{2026}
\acmJournal{PACMPL}
\acmVolume{10}
\acmNumber{OOPSLA2}
\acmArticle{387}
\acmMonth{10}
\acmSubmissionID{oopslab26main-p1146-p}
\received{2026-03-17}
\received[accepted]{2026-06-10}

\usepackage{./sty/prelude}

\begin{document}

\title{A Design Space Exploration of Async/Await}

\author{Gavin Gray}
\correspondingauthor
\orcid{0000-0002-2960-1198}
\affiliation{%
  \institution{Brown University}
  \city{Providence}
  \country{USA}
}
\email{gavin\_gray@brown.edu}

\author{Shriram Krishnamurthi}
\orcid{0000-0001-5184-1975}
\affiliation{%
  \institution{Brown University}
  \city{Providence}
  \country{USA}
}
\email{shriram@brown.edu}

\author{Will Crichton}
\orcid{0000-0001-8639-6541}
\affiliation{%
  \institution{Brown University}
  \city{Providence}
  \country{USA}
}
\email{will\_crichton@brown.edu}

\setlength{\enumerateparindent}{\parindent}

\begin{abstract}
  Many modern programming languages include some form of asynchronous programming. In particular, a growing number now have what we call straight-line asynchrony: attempts to provide asynchronous functions that look similar to synchronous functions, thereby enabling asynchrony without introducing complex control. These languages often share construct names like ``async'' and ``await,'' which suggests that they have deep semantic similarities. Yet, a close examination reveals that these languages are quite different along several dimensions, often subtly. These differences have real semantic consequences: similar-looking programs can exhibit divergent behavior, confusing developers and language designers alike.

  This paper therefore presents a \emph{design space exploration} of straight-line asynchrony. We dissect several existing languages, and show how no two of them agree as a whole on design decisions that affect the presence and ordering of execution. We articulate a design space with nine dimensions covering the full lifecycle of an asynchronous computation, covering questions such as: What precise guarantees does a language give upon calling an asynchronous function? What happens at the end of a task's life? How can a task handle being cancelled? We explore these questions through concrete examples, informal design discussion, and a formal semantics. Our ultimate goal is to help programmers, language designers, and language theorists all better understand the emerging landscape of straight-line asynchrony.
\end{abstract}



\maketitle

\section{Introduction}

Software developers have long struggled to write concurrent and parallel programs that are both correct and efficient. Programming languages have responded to this challenge by gradually lifting elements of concurrency from user-space to the level of language design. Approaches vary greatly, but one core idea shared between most modern languages is \emph{async/await}: marking blocks of code as \code|async|, and using an \code|await| keyword to wait for an async block to finish executing. These features can be found in recent versions of Python, JavaScript, Swift, Rust, \CSharp, and more.

Our original motivation for studying async/await was to help developers translate their understanding of the concept from one language to another, e.g., from JavaScript to Rust. In theory, both languages contain the same core concept with small differences in the details. We should be able to explain async/await just as one might explain lexically-scoped variables, if-statements, or other standardized language features~\cite{lu_identifying_2024}. In reality, the surface similarities between async/await designs can mask deep semantic divergences. For instance: what's the difference between a coroutine, a future, a promise, and a task? What happens when you cancel a task? When are tasks required to exit, if ever? How do errors propagate across a task boundary, if at all? These questions are further complicated by languages with customizable async/await runtimes (e.g., Rust and Python), which give different answers based on the choice of runtime.

\begin{figure}[t]
  \begin{subfigure}[t]{0.33\linewidth}
    \begin{lstlisting}[language=Pseudo]
async fn write_to_log():
  print("A")
  // simulate log write
  await sleep(2)
  print("B")

async fn process_await():
  task = spawn write_to_log()
  await sleep(0)
  await task

async fn process_detach():
  task = spawn write_to_log()
\end{lstlisting}
  \caption{Two examples of functions that spawn a task to write data to a log, one that waits for the task, and another that doesn't.}
  \label{fig:motivating-example-process}
  \end{subfigure}
  \hfill
  \begin{subfigure}[t]{0.28\linewidth}
    \begin{lstlisting}[language=Pseudo]
async fn ex1():
  t = write_to_log()
  print("C")
  await sleep(3)

async fn ex2():
  await process_detach()
  await sleep(1)
  print("C")

async fn ex3():
  await timeout(
    1, process_await)
  print("C")
  await sleep(3)
\end{lstlisting}
\cprotect\caption{Three example invocations of the processing functions.}
  \label{fig:motivating-example-invoke}
  \end{subfigure}
  \hfill
  \begin{subfigure}[t]{0.35\linewidth}
    \vspace{0pt}
    \begin{tabular}{lcccc}
    \toprule
                 & \code|ex1| & \code|ex2| & \code|ex3| \\
    \midrule
    \CSharp    & ACB & AC  & ACB \\
    \Js        & ACB & ACB & ACB \\
    Swift      & CAB & AC  & AC  \\
    Python \\
    ~~Asyncio    & C   & AC  & AC  \\
    ~~Trio       & C   & ABC & AC  \\
    Rust \\
    ~~Tokio      & C   & AC  & ACB \\
    ~~Smol       & C   & C   & AC  \\
    \bottomrule
    \end{tabular}
    \caption{A table of outcomes when running the programs translated from the pseudocode into different languages.}
    \label{fig:motivating-example-outcomes}
  \end{subfigure}
  \caption{An example of an asynchronous programming pattern that exhibits different behavior under seven modern async/await languages and runtimes. We discuss the translation from pseudocode to each respective language in \Cref{sec:designdims}.}
  \label{fig:motivating-example}
\end{figure}

As a concrete example, consider the pseudocode in \Cref{fig:motivating-example}, showing two ways to run a background task that writes information to a log.  Languages differ with respect to what happens when an async function is called, they differ with tasks at the end of their spawning scope, and languages also differ in terms of when and how tasks are stopped in the event of cancellation. When comparing seven modern approaches to async/await, \emph{no} two of them exhibit the same behavior on three straightforward calling contexts!

Observing this conceptual chaos, we set out to perform a \emph{design space exploration} of async/await in modern languages. The contribution of this paper is the result of that exploration:
\begin{enumerate}
  \item \textbf{We start by situating the work with a high-level overview of asynchronous programming} (\Cref{sec:motivation}). We contrast asynchrony against concurrency and parallelism, and introduce common asynchrony mechanisms such as racing and cancellation.

  \item \textbf{We describe the core design dimensions of asynchrony in modern languages} (\Cref{sec:designdims}). Our goal is to capture the most fundamental distinctions between designs, focusing on design choices that cause similar-looking programs to exhibit divergent behavior.

  \item \textbf{We provide a precise description of each design dimension using operational semantics} (\Cref{sec:semantics}). We give a precise model of each language's async design in terms of variants on a core asynchrony calculus with delimited continuations.
\end{enumerate}

\section{Core Concepts in Asynchrony}\label{sec:motivation}

Async/await are just keywords---what is the underlying concept being expressed across these languages? The obvious answer is \emph{asynchronous programming} or, more simply, \emph{asynchrony}.
Like the words concurrency and parallelism, asynchrony does not have a definitive meaning. Rather, its meaning is socially constructed out of the types of programs people want to write. Fifteen years ago, asynchronous programming was ``characterized by many simultaneously pending reactions to internal or external events''~\cite{syme2011async}. Today, we might add to that list: running \abbr{i/o}-bound computations in parallel, or having fine-grained control over the execution of computations. The scope of asynchronous programming has expanded.

Moreover, async/await is only one form of asynchrony. The job of async/await is to avoid the inversion of control associated with callbacks and event loops. The goal, as the \CSharp documentation says, is ``to enable code that reads like a sequence of statements, but executes in a more complicated order''~\cite{wagner_asynchronous_2025}. Async/await is about leaning on the language to simplify control-flow patterns, as opposed to implementing asynchrony abstractions entirely at the library level.

Before diving into the details of async design dimensions, we start by providing a high-level picture of async/await circa 2026: how does asynchrony relate to the similar concepts of concurrency and parallelism, and what does it look like to write async/await code today?

\subsection{Asynchrony vs.\ Concurrency vs.\ Parallelism}
\label{s:sync-quotes-from-langs}

We will define concurrency as a logical property of a computation, whereby two subcomputations can execute interleaved: one computation begins before another ends. Parallelism is a physical property of a computation, whereby two subcomputations can execute simultaneously: both computations overlap at the same point in time. Parallelism implies concurrency but not vice versa. For instance, an embedded operating system running on a single core is concurrent but not parallel.

We think of asynchrony as a special case of cooperative concurrency. In an asynchronous program, two subcomputations execute interleaved because they are written to yield control to one another. Asynchrony contrasts with competitive concurrency, where interleaving occurs through external imposition, e.g., by context switches. Competitive concurrency usually arises at an architectural level below the language, such as operating system threads and interrupts.

Concurrency and parallelism have programming paradigms. Paradigms of concurrent computing include shared-memory concurrency (e.g., pthreads), communicating sequential processes (MPI, Go), and actors (Erlang). Paradigms of parallel computing include fork-join parallelism (Cilk, OpenMP), model parallelism (MapReduce, Spark), and data parallelism (OpenGL, CUDA).  Like all paradigms, these labels are useful but leaky, such as how a complex parallel program often involves shared-memory concurrency. Ultimately, a paradigm is about focusing a set of related mechanisms (\code|spawn|, mutexes) on a set of related problems (sorting a list, running a database).

Asynchrony also has paradigms. Callback asynchrony uses higher-order functions to provide a continuation for the completion of a computation, most notably found in \CSharp{}, \Js, and Swift. Event-loop asynchrony structures a program around a main loop that dispatches reactions to an event stream, most notably found in \Cpp and Rust. Languages have historically developed async/await to address the ergonomic problems associated with these two asynchrony paradigms. A key motivation is to provide a surface syntax for concurrent code that looks comparable to non-concurrent (synchronous) code. For example, below are excerpts from async/await proposal documents across a few languages (see the Supplemental Material for additional examples):

\begin{displayquote}
  \textbf{(\CSharp{})} ``Notice first how similar it is to the synchronous code. \ldots The control flow is completely unaltered, and there are no callbacks in sight. That doesn’t mean that there are no callbacks, but the compiler takes care of creating and signing them up \ldots''~\cite{torgersen_asynchronous_nodate}
\end{displayquote}

\begin{displayquote}
  \textbf{(Python)} ``It is proposed to make coroutines a proper standalone concept in Python, and introduce new supporting syntax. The ultimate goal is to help establish a common, easily approachable, mental model of asynchronous programming in Python and make it as close to synchronous programming as possible.''~\cite{selivanov_pep_2015}
\end{displayquote}

\begin{displayquote}
  \textbf{(Rust)}
  ``From a user's perspective, they can use async/await as if it were synchronous code, and only need to annotate their functions and calls.''~\cite{withoutboats_2394_2018}
\end{displayquote}

\subsection{Straight-Line Asynchrony}
\label{sec:straightline-async}

We therefore define the paradigm embodied by async/await as \emph{straight-line asynchrony}, given its fundamental organizing principle of eliminating complex control flow. An alternative name would be \emph{task asynchrony}, used within the \CSharp community. As we will discuss, tasks are indeed a central concept to async/await, but when looking cross-linguistically, tasks are not the only element of straight-line asynchrony. For instance, languages like Rust provide coroutines so straight-line asynchrony can be achieved without ever creating a task. Some designs like Python+Trio do not expose any concept of a task to the user.

The goal of this paper is to explore the design space of straight-line asynchrony. We primarily focus on programming languages that (a) have a relatively settled design for straight-line asynchrony, and (b) are in relatively widespread use. That includes languages with a singular form of asynchrony: \Js (specifically ECMAScript 2025, with TypeScript types for clarity), \CSharp (v14 on .NET~10), and the \emph{structured-concurrency} subset of Swift (v6.2). It also includes languages with a modular form of asynchrony: Python (v3.14, using Asyncio and Trio v0.33) and Rust (v1.92.0, with Tokio v1.50.0 and \smol v2.0.2). We discuss how our findings relate to other languages in \Cref{sec:discussion}.
Throughout, we consider only async function calls that introduce concurrency. We do not consider immediately-awaited applications (e.g., \code|await f()|), which complete before the caller resumes and therefore behave like synchronous calls under every design we study.
We describe our findings informally in \Cref{sec:designdims} and formally in \Cref{sec:semantics}.

We do not expect our readers to be familiar with the concrete syntax of all relevant languages, so we articulate the design dimensions at a high level, and present code examples primarily using the same pseudocode style in \Cref{fig:motivating-example}. However, it is useful to first get a sense for both the gestalt and the variety of syntaxes, mechanisms, and semantics in straight-line asynchrony. We therefore conclude this section by implementing an async utility in multiple languages.

\begin{figure}[t]
  \begin{subfigure}{0.33\linewidth}
\begin{lstlisting}[language=Js]
async function timeout<T>(
  secs: number,
  f: () => Promise<T>
): Promise<T | null> {
  return await Promise.race([
    f(), sleep(secs)
  ]);
}
\end{lstlisting}
\cprotect\caption{\Js using its built-in Promise \abbr{api}. \code|Promise.race| returns the result of whichever future finishes first. \Js does not cancel the losing computation.}
  \end{subfigure}
  \hfill
  \begin{subfigure}{0.30\linewidth}
  \begin{lstlisting}[language=Py]
async def timeout(
  secs: float,
  f: Callable[[],
       Awaitable[T]]
) -> T | None:
  with trio.move_on_after(
    secs
  ):
    return await f()
\end{lstlisting}
  \cprotect\caption{Python+Trio using its first-class construct for timeout. Unlike other languages sampled, Python+Trio does not provide a racing primitive. The timed-out computation is cancelled.}
  \label{fig:timeout-simple-trio}
\end{subfigure}
  \hfill%
  \begin{subfigure}{0.32\linewidth}
  \begin{lstlisting}[language=Rust]
async fn timeout<F: Future>(
  d: Duration,
  f: impl FnOnce() -> F
) -> Option<F::Output> {
  tokio::select! {
    result = f() =>
      Some(result),
    _ = sleep(d) =>
      None,
  }
}
\end{lstlisting}
  \cprotect\caption{Rust+Tokio using \code|select!|, which dispatches based on the result of a race. The losing computation is cancelled.}
  \end{subfigure}
\cprotect\caption{A simplified implementation of \code|timeout|, demonstrating primitives for racing. The minimal semantics for timing out does not guarantee that a timed-out computation is cancelled, it simply guarantees that execution is no longer blocked.}
\label{fig:timeout-simple}
\end{figure}

\subsection{A Case Study on Timing Out}
\label{sec:timeout}

Consider implementing a \code|timeout| function, which takes as input an asynchronous function and a duration. The minimal semantics is that \code|timeout| returns either the result of the asynchronous work, or if the work takes longer than the duration, an indication of timing out. The \code|timeout| function is a paradigmatic example of asynchronous programming, as its implementation is often non-trivial when using only primitives from other concurrency or parallelism paradigms.

Most straight-line async designs include a race primitive (also called \code|select|, \code|or|, \code|wait|, or \code|next|) that returns the first of several asynchronous computations to complete. \Cref{fig:timeout-simple} shows three implementations of the minimal \code|timeout| semantics using these primitives. The dual of racing is to wait for all computations to complete, variously called \code|all|, \code|and|, \code|join|, or \code|gather|.

An enhanced semantics for \code|timeout| should take into account \emph{cancellation}: stopping a computation that is no longer needed. Cancellation is another core concept for asynchrony, which did not feature prominently in earlier designs such as \Js and \CSharp, but is more consciously part of later designs such as Swift and Python. For \code|timeout|, we desire three properties. First, if \code|f| exceeds the duration, then \code|f| is cancelled. Second, if \code|timeout| is cancelled, then \code|f| is also cancelled. Third, if \code|f| is cancelled, then it can gracefully shut down, i.e., execute cleanup code prior to termination.

\begin{figure}[t]
\begin{subfigure}{0.51\linewidth}
  \begin{lstlisting}[language=CSharp]
async Task<T> Timeout<T>(
  TimeSpan duration,
  CancellationToken token,
  Func<CancellationToken, Task<T>> f
) {
  using var childCts = CancellationTokenSource
    .CreateLinkedTokenSource(token);
  var work = f(childCts.Token);
  try {
    return await work.WaitAsync(
      duration, token);
  } catch (Exception e) when (
    e is TimeoutException or OperationCanceledException
  ) {
    childCts.Cancel();
    throw;
  }
}
\end{lstlisting}
\cprotect\caption{\CSharp does not provide a built-in mechanism to cancel a \code{Task}, graceful shutdown requires the use of a cancellation token threaded through each async function. The use of a child token, \code|childCts|, ensures that cancelling \code|f| does not cancel other tasks observing the \code|token| input to \code|Timeout|.}
\label{fig:timeout-tokio}
\end{subfigure}%
\hfill
\begin{subfigure}{0.47\linewidth}
  \begin{lstlisting}[language=Swift]
func timeout<T: Sendable>(
  duration: Duration,
  f: @Sendable () async throws -> T
) async throws -> T? {
  try await withoutActuallyEscaping(f) {
    f in
    try await withThrowingTaskGroup {
      group in
      group.addTask { try await f() }
      group.addTask {
        try await Task.sleep(for: duration)
        return nil
      }
      defer { group.cancelAll() }
      return try await group.next()!
    }
  }
}
\end{lstlisting}
\cprotect\caption{Swift. The \code|TaskGroup| interface automatically enforces that \code|f| is cancelled if \code|timeout| is cancelled, and the \code|defer| block enforces that \code|f| is cancelled if the sleep returns first. Each block is wrapped in \code|try| because awaiting a cancelled task can raise an exception.}
\end{subfigure}
  \cprotect\caption{Enhanced implementations of \code|timeout| handling cancellation both from within and without.}
  \label{fig:timeout-cancel}
\end{figure}

These goals can be achieved either through library-level or language-level implementations of cancellation. At the library level, a token or signal is threaded through a program, and each task is responsible for observing that token. At the language level, tasks are grouped together, and cancelling one task propagates the cancellation to other tasks, often by throwing a cancellation exception. \Cref{fig:timeout-cancel} shows a library-level and language-level implementation of the enhanced \code|timeout| semantics.

For the pseudocode in this paper we use the minimal semantics implementation of \code|timeout| as our cancellation primitive. We made this choice to encompass the varying primitives exposed by runtimes, as well as to accommodate languages without built-in cancellation (e.g., \Js, \CSharp).

In sum, \Cref{fig:motivating-example} demonstrated how different designs for straight-line asynchrony can take the same simple-looking program and cause different outcomes. This section demonstrated how for more sophisticated asynchronous patterns, the same concept can require substantively different implementations. Together, they motivate the need for a better understanding of the collective design space of straight-line asynchrony.

\newcommand{\DimSection}[1]{%
  \multicolumn{4}{@{}l}{\large #1} \\ \addlinespace[1ex]
}

\newcommand{\DimRowTitle}[2]{%
  #1 \newline {\footnotesize #2\par}
}

\newcommand{\DimCell}[3]{%
  \textsc{#1} \newline {\footnotesize #2\par\vspace{0.5ex}{\emph{#3}}\par}
}

\newcommand{\DimSeparator}{%
  \arrayrulecolor{gray}\specialrule{0.5pt}{1ex}{1.5ex}\arrayrulecolor{black}
}

\newcommand{\LangSep}{~$\cdot$\space}

\begin{table}[htbp]
\centering
\setlength{\parskip}{0pt}
\begin{tabularx}{\textwidth}{@{}
  >{\raggedright\arraybackslash}p{3.2cm}
  >{\raggedright\arraybackslash}X
  >{\raggedright\arraybackslash}X
  >{\raggedright\arraybackslash}X
  @{}}

\DimSection{Start of Life}

\DimRowTitle{Eagerness}{How to evaluate an async function application.} &
\DimCell{\lazy}{Evaluate to a coroutine without executing further.}{Python\LangSep Rust} &
\DimCell{\eager}{Evaluate in current thread, and schedule as task on await.}{\CSharp\LangSep \Js\LangSep Swift~\emph{(immediate)}} &
\DimCell{\seager}{Evaluate to task, and schedule task for execution.}{Swift~\emph{(async let)}} \\
\addlinespace[1ex]

\DimRowTitle{Suspension}{Guarantees on whether await points suspend.} &
\DimCell{\static}{Await points guaranteed to suspend.}{\Js} &
\DimCell{\dynamic}{No guarantees on awaiting tasks.}{\CSharp\LangSep Swift\LangSep Tokio\LangSep Smol\LangSep Asyncio\LangSep Trio} & \\

\DimSeparator

\DimSection{End of Life}

\DimRowTitle{Extent}{The default interval of time during which a task may exist.} &
\DimCell{\indef}{Tasks by default may exist until the end of the runtime.}{\Js\LangSep \CSharp\LangSep Tokio\LangSep Smol\LangSep Asyncio} &
\DimCell{\dyn}{Tasks by default may exist until the end of their spawning scope.}{Swift\LangSep Trio} & \\
\addlinespace[1ex]

\DimRowTitle{Reference Strength}{[For \indef \extent{}] \newline The type of reference to a task held by the runtime.} &
\DimCell{\strong}{The runtime holds a strong reference.}{\Js\LangSep \CSharp\LangSep Tokio} &
\DimCell{\weak}{The runtime holds a weak reference.}{Asyncio\LangSep Smol} & \\
\addlinespace[1ex]

\DimRowTitle{Destruction}{How a task is cleaned up at the end of its extent.} &
\DimCell{\awaited}{The task is awaited to completion.}{\Js\LangSep Trio} &
\DimCell{\cancelled}{The task is cancelled, and then possibly awaited.}{Swift\LangSep Tokio\LangSep Smol\LangSep Asyncio} &
\DimCell{\terminated}{The program exits.}{\CSharp} \\
\addlinespace[1ex]

\DimRowTitle{Propagation}{What happens to exceptions in unawaited tasks.} &
\DimCell{\destruct}{Exceptions are reraised by dependents.}{Trio} &
\DimCell{\never}{The exception is kept within the task.}{\Js\LangSep \CSharp\LangSep Tokio\LangSep Smol\LangSep Asyncio\LangSep Swift} & \\

\DimSeparator

\DimSection{Cancellation}

\DimRowTitle{Awareness}{Whether a task is able to respond to being cancelled.} &
\DimCell{\unaware}{The task cannot respond to being cancelled.}{Rust} &
\DimCell{\aware}{The task can respond to being cancelled.}{Asyncio\LangSep Trio\LangSep Swift} & \\
\addlinespace[1ex]

\DimRowTitle{Direction}{How cancellation is communicated through the task graph.} &
\DimCell{\tdown}{Starting from the root task, and communicated from dependents to dependencies.}{Rust} &
\DimCell{\bup}{Starting from the root's dependencies and communicated to dependents.}{Asyncio\LangSep Trio} &
\DimCell{\simultaneous}{To all transitive dependencies at once.}{Swift} \\
\addlinespace[1ex]

\DimRowTitle{Persistence}{[For \aware cancellation] \newline How long a cancellation of a task lasts.} &
\DimCell{\transient}{A task can ignore cancellation and proceed as normal.}{Asyncio} &
\DimCell{\persistent}{A task can ignore cancellation but remains cancelled.}{Trio\LangSep Swift} & \\

\end{tabularx}
\caption{Overview of the asynchronous programming design dimensions}
\label{tab:dims-overview}
\end{table}

\section{Design Dimensions}\label{sec:designdims}

Every design for straight-line asynchrony is composed of dozens of individual design decisions that trade off performance, correctness, usability, extensibility, interoperability, and so on. To scope this paper, we focus on design decisions that substantively influence the \emph{functionality} of asynchronous programs. As suggested by \Cref{fig:motivating-example}, we want to explain how similar-looking programs can have divergent behavior in terms of the order or presence of operations.

By contrast, this paper does not attempt to account for design decisions that are principally about \emph{ergonomics}: for example, should \code|await| be a prefix or postfix operator in the concrete syntax?
This paper also does not focus on design decisions that are principally about \emph{performance}: for example, how should an asynchronous runtime efficiently interface with the operating system? However, for straight-line asynchrony, performance and functionality are difficult to disentangle. Several design dimensions we discuss will be motivated by performance, but the key criterion for inclusion is their ultimate impact on functionality.


To perform this design space exploration, we manually analyzed the designs of today's languages containing straight-line asynchrony features. We ultimately focused on the languages enumerated in \Cref{sec:straightline-async} (\Js, \CSharp, Swift, Python, Rust), but we also examined Kotlin, \Cpp{}20, \FSharp, Haskell, OCaml, Hack, Nim, Dart, and Zig. We read design documents (e.g., RFCs, published papers), GitHub discussions, and community blogs. In addition, we drew on reports from community members of confusing or surprising behavior in async programs~\cite{saad_using_2021,william_how_2019,lacuna_asyncio_nodate,smith_thoughts_2016,paharia_cancelling_2025,paharia_400_2025,pacheco_397_nodate,smith_notes_2018,archibald_gotcha_2023}.


We identified nine design dimensions, collected into \Cref{tab:dims-overview}. These dimensions are grouped into three categories, \StartOfLife, \EndOfLife, and \Cancellation, which serve to organize the remainder of this section. For each design dimension, we explain the points along that dimension, and the design justifications offered by different languages for selecting each point.

\subsection{\StartOfLife}

An asynchronous computation is generally defined through a function explicitly marked as asynchronous, such as the \lstinline[language=Pseudo]|async fn write_to_log| in \Cref{fig:motivating-example-process}. To begin invoking an asynchronous function in any language, one simply calls the function, as in \code|write_to_log()|. Immediately, the semantics diverge along the first axis: \Eagerness.

\subsubsection{\Eagerness}
\label{ssec:designdims:eagerness}

\begin{figure}[t]
  \centering
  \includegraphics[width=0.82\linewidth]{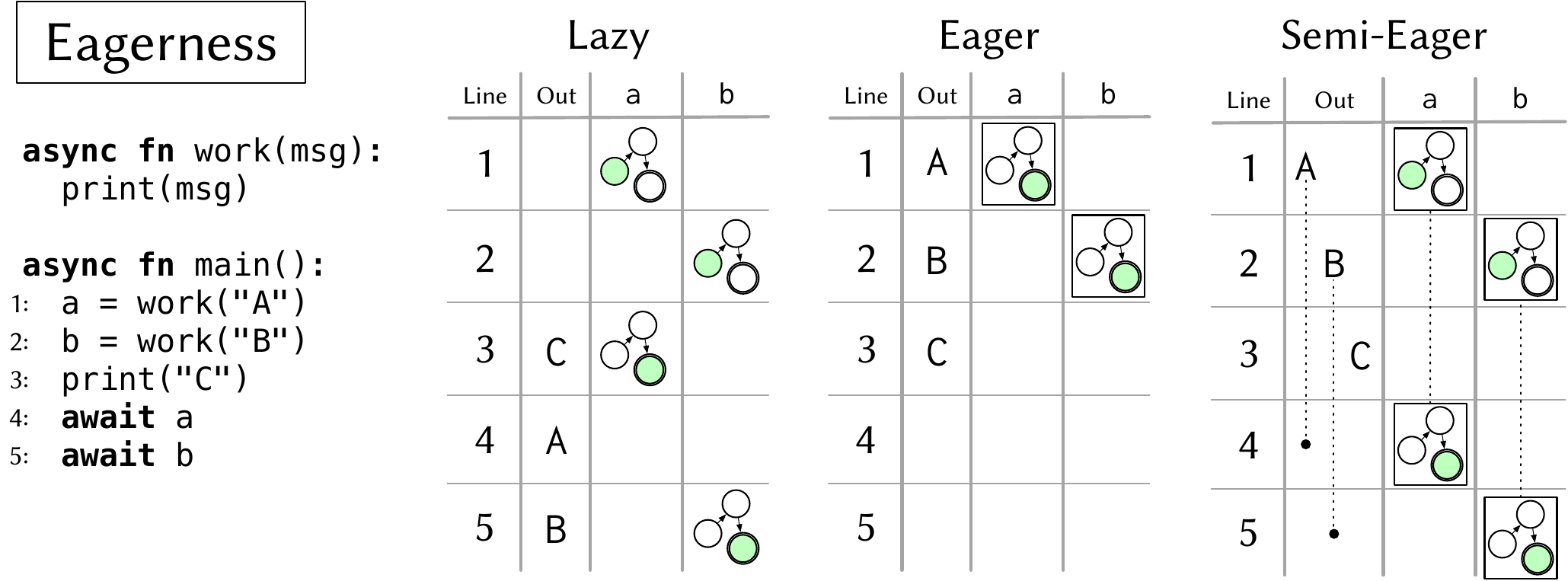}
  \cprotect\caption{An example program demonstrating the \eagerness dimension, executed with Rust (\lazy), \CSharp (\eager), and Swift (\seager) when using \emph{async let.} Each column shows the line of execution, the text printed to stdout, and changes to the values of the variables \code|a| and \code|b|. Coroutines are represented as a state machine, with the active state colored in green. Tasks are represented as a boxed state machine. The dotted vertical lines represent nondeterminism, or an event that could happen anywhere within the indicated range.}
  \label{fig:eagerness}
\end{figure}

Consider the program in \Cref{fig:eagerness}, where an async function \code|work| is called twice, and later awaited. Based on each language's approach to async function calls, three outcomes can occur: \lazy, where no work occurs until being awaited, \eager, where work occurs immediately in the main thread of execution, and \seager, where no work occurs but is immediately registered in the runtime to start soon, potentially on a different thread, resulting in non-deterministic output. To understand this variation, we first need to discuss two foundational concepts for straight-line asynchrony: coroutines and tasks.

An asynchronous function either implicitly (in the runtime implementation) or explicitly (in a user-visible way) desugars into an object that behaves like a coroutine. Upon entering the function, code runs until reaching an \code|await| point, when the function may yield to its caller (depending on its \Suspension, \Cref{ssec:designdims:suspension}). Specifically, these objects behave like asymmetric stackless coroutines, to use the taxonomy provided by \citet{moura_revisiting_2009}. General coroutines can yield values on suspension and accept values on resumption. For straight-line asynchrony, a coroutine does not need to yield values on suspension, but needs only to accept values on resumption (or access them at a shared location).

When a coroutine representing an asynchronous computation is constructed via a function call, the \eagerness axis defines what happens next to the coroutine. In languages with first-class coroutines\footnotemark{} such as Python~\cite{selivanov_pep_2015} and Rust~\cite{withoutboats_2394_2018}, we describe their behavior as \lazy: the coroutine is returned and no work occurs. In \Cref{fig:eagerness}, this means that \code|a = work("A")| constructs a coroutine in a starting state, but the coroutine is only executed at \code|await a| on line 4. In \lazy languages, to execute an async computation concurrently with its caller, one must use an explicit \lstinline[language=Pseudo]|spawn| operation to convert the coroutine into a \emph{task} (e.g., as used in \Cref{fig:motivating-example}).
\footnotetext{As of v1.92.0, Rust has first-class \emph{futures} but not first-class coroutines. That is, calling an async function returns a value representing the computation, but that value does not offer a full coroutine interface. By contrast, Python has a stabilized coroutine interface implemented by its async functions. The Rust compiler does internally track async functions as coroutines, but this interface has not been stabilized.}

A task is a handle on an async computation shared between its creator and an \emph{async runtime}. The async runtime usually consists of a thread pool and some facilities for reacting to \abbr{os}-level events by waking up waiting tasks. In languages without first-class coroutines, a call to an async function will return a task directly. However, languages differ with respect to scheduling of these tasks.

For \eager languages such as \CSharp~\cite[\textsection 15.14.3]{ecma_c_standard_committee_c_nodate} and \Js~\cite[\textsection 27.7.5.1]{ecma_ecmascript_nodate}, on evaluating an async function application, the current thread transfers execution to the body of the async function. When reaching an await point, the function application evaluates to a \task value, and control returns to the caller. In \Cref{fig:eagerness}, \eager execution causes \code|work("A")| to behave like a normal function call, resulting in deterministic/single-threaded execution.

A language is \seager when an async function application immediately schedules the task onto a work queue, but allows the calling thread to continue executing past the function call. In \Cref{fig:eagerness}, \seager execution means that \code|work("A")| will run at some point after being called. The task is only guaranteed to have finished by the end of \code|await a|, leading to nondeterministic/multi-threaded execution. We illustrate the \seager design using Swift~\cite{mccall_swift_2021-1} by assuming \emph{async let}~\cite{apple:async-let} bound async applications.\footnotemark{} Unlike all other languages surveyed, a bare async application in Swift (e.g., \code|let a = work("A")|) results in a type error. Async applications must be immediately awaited and developers can use alternative syntaxes to run an async function within a new \task. We classify Swift as being both \eager and \seager as it provides alternative syntax for both semantics and doesn't itself have a default. For consistency, all pseudocode in this paper assumes that async function applications in Swift are bound using \emph{async let}.%
\cprotect\footnotetext{Although we illustrate the \seager design using Swift, Swift is not \seager by default. To call an async function developers must choose between an immediately-awaited application (e.g., \code|await work("A")|), a \seager application (e.g., \code|async let x = work("A")|), or an \eager application (e.g., \code|Task.immediate { await work("A") }|). A bare async application with no specification (e.g., \code|let x = work("A")|) is disallowed in Swift and results in a type error.}

\paragraph{Design rationale.} The choices in the \eagerness dimension offer a tradeoff in predictability, maximizing \abbr{cpu} multi-core utilization, and reducing \task allocation.
\Lazy asynchrony has two main benefits. First, \lazy asynchrony can provide predictable memory allocation. Asynchronous function applications evaluate directly to a \coro value that can be stack allocated. The developer must opt in to either boxing or spawning a \coro. Predictable memory allocation is a key motivation for Rust's async/await design~\cite{withoutboats_why_2023}.
Second, by separating coroutines from tasks, the language can enable users to provide their own runtime. For example in Rust, this feature is useful so the runtime can be tailored to the needs of individual domains, such as embedded systems running Embassy~\cite{rust:embassy} versus backend web servers running Tokio/Smol.

The downsides of the \lazy approach include: by offering an extensible runtime, async-related libraries in the ecosystem tend to have reduced interoperability. For example, in Rust, \lstinline[language=Pseudo]|spawn| is a runtime-specific primitive, so libraries using it must be coupled to a choice of runtime such as Tokio or Smol. Users must also deal with the complexity of understanding both coroutines and tasks. Users often find \lazy behavior surprising---to call a function and have nothing happen runs against a programmer's intuition transferred from synchronous function calls~\cite{d_why_2025,gebra_how_2023,mapeper_starting_2023}.

Non-\lazy approaches avoid these downsides by starting asynchronous work as soon as a function is called, either on the current thread (\eager) or any thread (\seager). The \eager approach reduces the overhead of control flow and context switches within the asynchrony runtime. The \seager approach leans more on the runtime, but in exchange provides a higher degree of concurrency by dispatching work to other threads. The increased concurrency can be particularly valuable for programs that want to avoid blocking a main thread, such as in \abbr{ui} applications~\cite{bartlett_why_2023,inc_embracing_nodate,winney_using_2021}.

\subsubsection{\Suspension}\label{ssec:designdims:suspension}

\begin{wrapfigure}{r}{0.52\linewidth}
  \includegraphics[width=\linewidth]{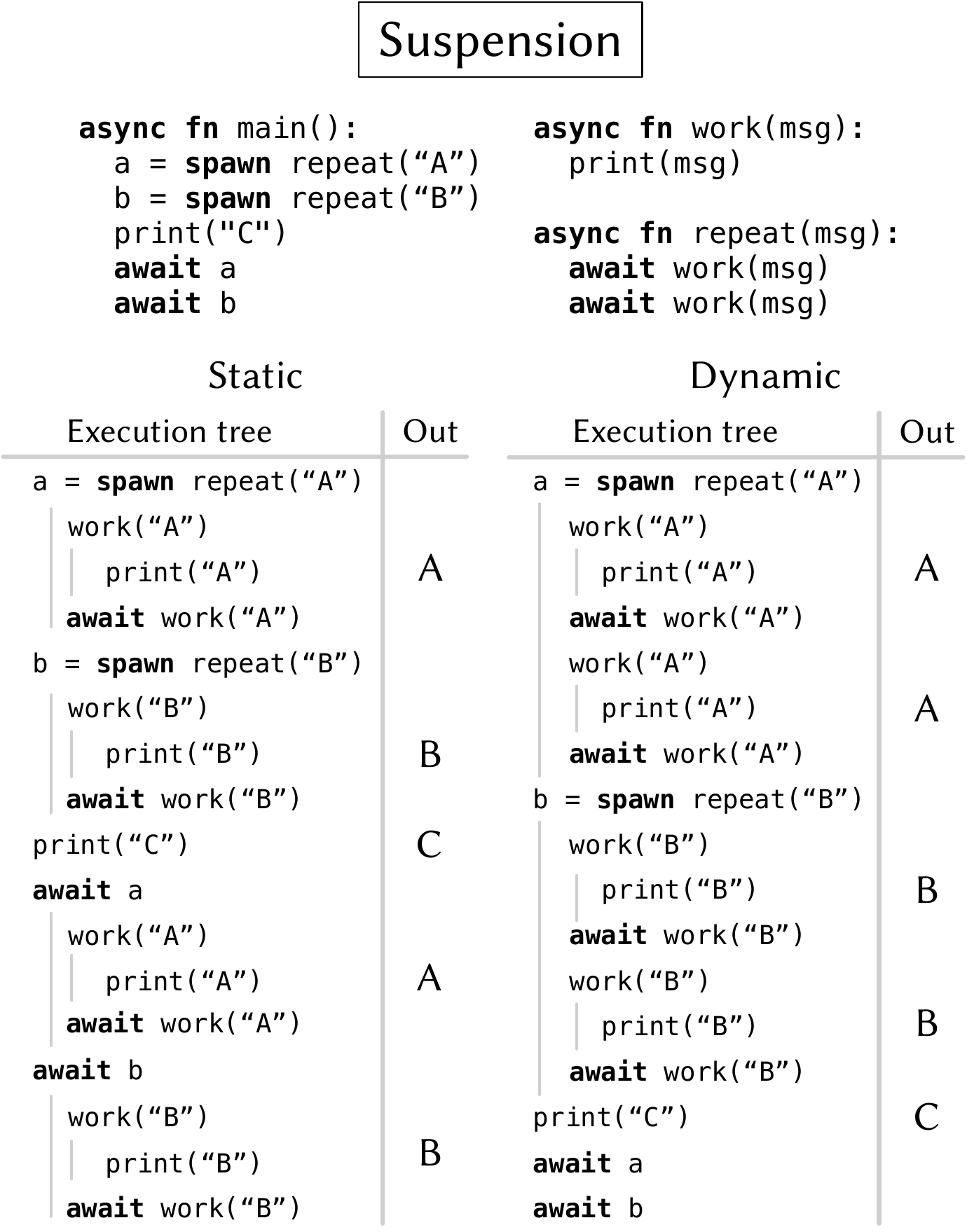}
  \caption{An example program demonstrating \Suspension, executed with \Js (\static) and \CSharp (\dynamic). Both languages are \eager, but have different \suspension semantics. Each column shows an execution trace and the order of prints to stdout; the traces are nested according to their level in the call stack.}
  \label{fig:suspension}
\end{wrapfigure}

Once asynchronous work has started, the next design divergence occurs at await points. The \suspension dimension defines the guarantees a language makes about whether an await point is guaranteed to yield control. For example, consider the program in \Cref{fig:suspension} executing under an \eager semantics. The line \code|a = spawn repeat("A")| will call into \code|repeat("A")|, which calls into \code|work("A")|, printing \code|"A"|, and returns to \code|await work("A")|. The task for \code|work("A")| has been completed, so a language has two choices: either yield control to the caller, or continue evaluation past the \code|await|. We call the former strategy \static \suspension because an await point is statically guaranteed to always yield. We call the latter strategy \dynamic \suspension because whether an await point yields depends on the awaited task.

In only one language surveyed, \Js, an await point is guaranteed to yield according to the ECMAScript specification \cite[\textsection 27.7.5.3, step 3.b]{ecma_ecmascript_nodate}. Therefore in \Cref{fig:suspension}, running under \Js semantics, the program outputs ``ABCAB'' because \code|await work("A")| yields back to \code|main|. In all other languages, no such guarantee exists. For the example program operating under an \eager semantics such as in \CSharp, the program would output ``AABBC'' because \code|await work("A")| observes that \code|work("A")| is complete and therefore the await expression continues evaluation.

\paragraph{Design rationale.}

The two principal reasons to adopt \static \suspension are preventing starvation and reducing non-determinism. In practice, some asynchronous function calls complete synchronously. For example, an asynchronous file read can return and buffer more bytes than requested; subsequent file reads within the buffered range return synchronously and perform no actual \abbr{i/o}. Under an \eager application semantics with \dynamic \suspension, a running \task can starve the others by never suspending. Although a task's code contains await points, it can execute entirely synchronously, turning an \abbr{i/o}-bound task into a \abbr{cpu}-bound task. A \static \suspension policy guarantees that tasks suspend at await points and don't accidentally starve others. Furthermore, \static \suspension is more predictable because it eliminates the non-deterministic factor of asynchronous functions executing synchronously.

\Static \suspension comes at a performance cost. In the case of awaiting a completed task, \static \suspension requires an expensive round-trip to the runtime to continue past the await. Languages with \dynamic \suspension can mitigate the starvation issue by providing a \code|yield_now| primitive that has no effect but to force a suspension to the runtime.

\subsection{\EndOfLife}

Much of the divergence between designs for straight-line asynchrony is about when and how tasks end. Recent designs often introduce some kind of ``structured'' asynchrony to ensure that tasks are accounted for at all times, and do not end up dangling or erroring unbeknownst to the rest of the program. This subsection covers the design dimensions of \extent (how long a task lives, \Cref{ssec:designdims:extent}), \references (the type of reference to tasks held by the runtime, \Cref{ssec:designdims:references}), \destruction (what happens to destructed tasks, \Cref{ssec:designdims:destruction}), and \propagation (what happens to unawaited exceptions, \Cref{ssec:designdims:propagation}).

\subsubsection{\Extent}\label{ssec:designdims:extent}

\begin{figure}[t]
  \includegraphics[width=0.82\linewidth]{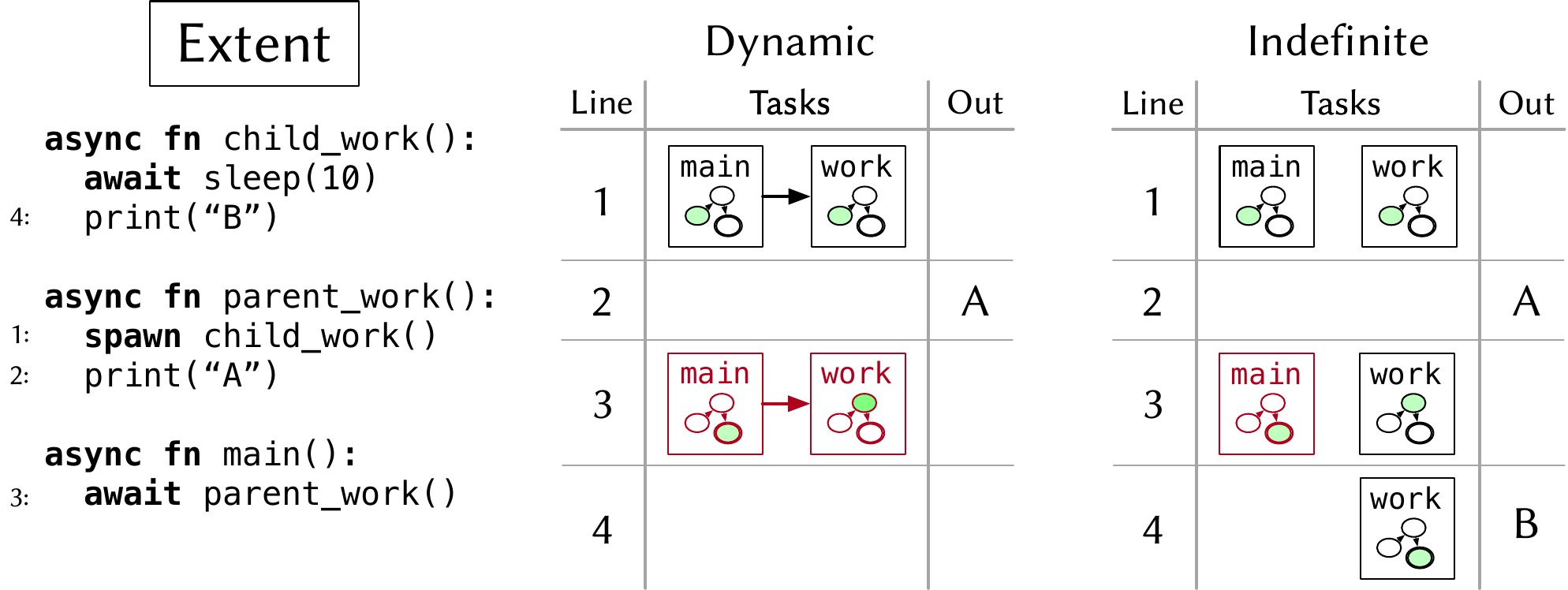}
  \caption{An example program demonstrating the \extent dimension, executed with Swift (\dyn) and \Js (\indef). Each column shows the task structure and output after executing each line. Arrows between tasks indicate a parent-child relationship used for \dyn \extent.}
  \label{fig:extent}
\end{figure}


Borrowing terminology from the Lisp community, the \extent of a task is the interval of time during which that task is capable of executing. For example, consider the program in \Cref{fig:extent}. What is the extent of the \code|spawn child_work()| task? The answer is it depends on quite a number of factors, as previously illustrated in \Cref{fig:motivating-example}.

In some designs for straight-line asynchrony, especially earlier designs, tasks by default have an \indef \extent: the task can live until it is no longer reachable, at which point the task is destructed. \Js, \CSharp, Python+Asyncio, and Rust+Tokio/Smol all use \indef \extent. Note that designs differ on how to determine reachability (see \Cref{ssec:designdims:references}), and designs also differ on the mechanism for destruction (see \Cref{ssec:designdims:destruction}). \Cref{fig:extent} illustrates \indef \extent, showing how the \code|child_work()| task outlives its parent \code|parent_work()| task, and the program waits until the work is completed.

Some have called \indef \extent an ``unstructured'' approach to asynchrony~\cite{smith_thoughts_2016}, akin to using goto for control flow, as task handles can easily go unhandled, leaving asynchronous work dangling with no clear policy for managing it.
This line of thought has led to the development of ``structured'' asynchrony approaches~\cite{sustrik_structured_2025,mccall_swift_2021,elizarov_kotlin_2021}, which we describe as having \dyn \extent, found in Swift and Python+Trio. With \dyn \extent, a task's extent is tied to the extent of another task. In \Cref{fig:extent}, under \dyn \extent, the \code|child_work()| task is associated with the \code|parent_work()| task. When the \code|parent_work()| task completes, its associated tasks are destructed, which means cancellation in the case of Swift. Therefore the second print never occurs.

\paragraph{Design rationale.}

The main argument for \dyn \extent is that limiting tasks to known scopes leads to programs that are easier to reason about, and compose better with other language features~\cite{mccall_swift_2021,elizarov_kotlin_2021}. For instance in Python, \dyn \extent can ensure that a task within a \code|with| block will not use resources (e.g., file descriptors) after they have been cleaned up~\cite{smith_thoughts_2016}. However, existing implementations of \dyn \extent only support a strictly hierarchical model of parent-child task relationships (via \code|TaskGroup| in Swift and \code|Nursery| in Trio). These implementations disallow programs that might have more complex dependencies between tasks, such as two parent tasks awaiting the same child task.

\subsubsection{\References}\label{ssec:designdims:references}

For designs with \indef \extent, tasks live until no longer reachable. One lever available to these designs is to ask: does the runtime's reference to a task count? In the parlance of reference counting garbage collection, is the runtime's reference \emph{strong} or \emph{weak}? This choice constitutes the \references dimension.

For most \indef \extent designs, the runtime's handle is \strong: \Js, \CSharp, Rust+Tokio. Consequently, even if a task is no longer accessible from the main body of a program, it is still reachable by the runtime, and will therefore be executed to completion. Two designs use \weak runtime handles: Rust+Smol and Python+Asyncio.

Rust lacks a garbage collector, so \weak \references does not translate literally for Smol. In Smol, the task handle received from a \lstinline[language=Pseudo]|spawn| implements a destructor (via the \code|Drop| trait), called when the handle goes out of scope (unless moved). The handle's destructor destructs the task, cancelling it in the case of Smol. In this way, Smol \emph{effectively} achieves \dyn \extent, unless the task is explicitly returned from the function. For example, the \dyn \extent semantics for the program in \Cref{fig:extent} are the same as Smol's semantics, and the \indef semantics are the same as Tokio's semantics.

Asyncio also holds \weak references to \tasks. Unlike Smol's intentional design, Asyncio's \weak reference approach is presently marked as a bug on the CPython repository, although the rationale for this design is unclear. According to Guido van Rossum~\cite{hartl_asyncio_2022}:

\begin{displayquote}
I'm not sure why we designed it this way, and (from private email) @1st1 doesn't seem to recall either. But it was definitely designed with some purpose in mind -- the code is quite complex and was updated every time an issue with it was discovered, and we've had many opportunities to change this behavior but instead chose to update the docs \citegithub{https://github.com/python/cpython/pull/29163}, even when asyncio itself suffered \citegithub{https://github.com/python/cpython/issues/90467}.~
\end{displayquote}

\subsubsection{\Destruction}\label{ssec:designdims:destruction}

\begin{figure}[tb]
  \includegraphics[width=0.8\linewidth]{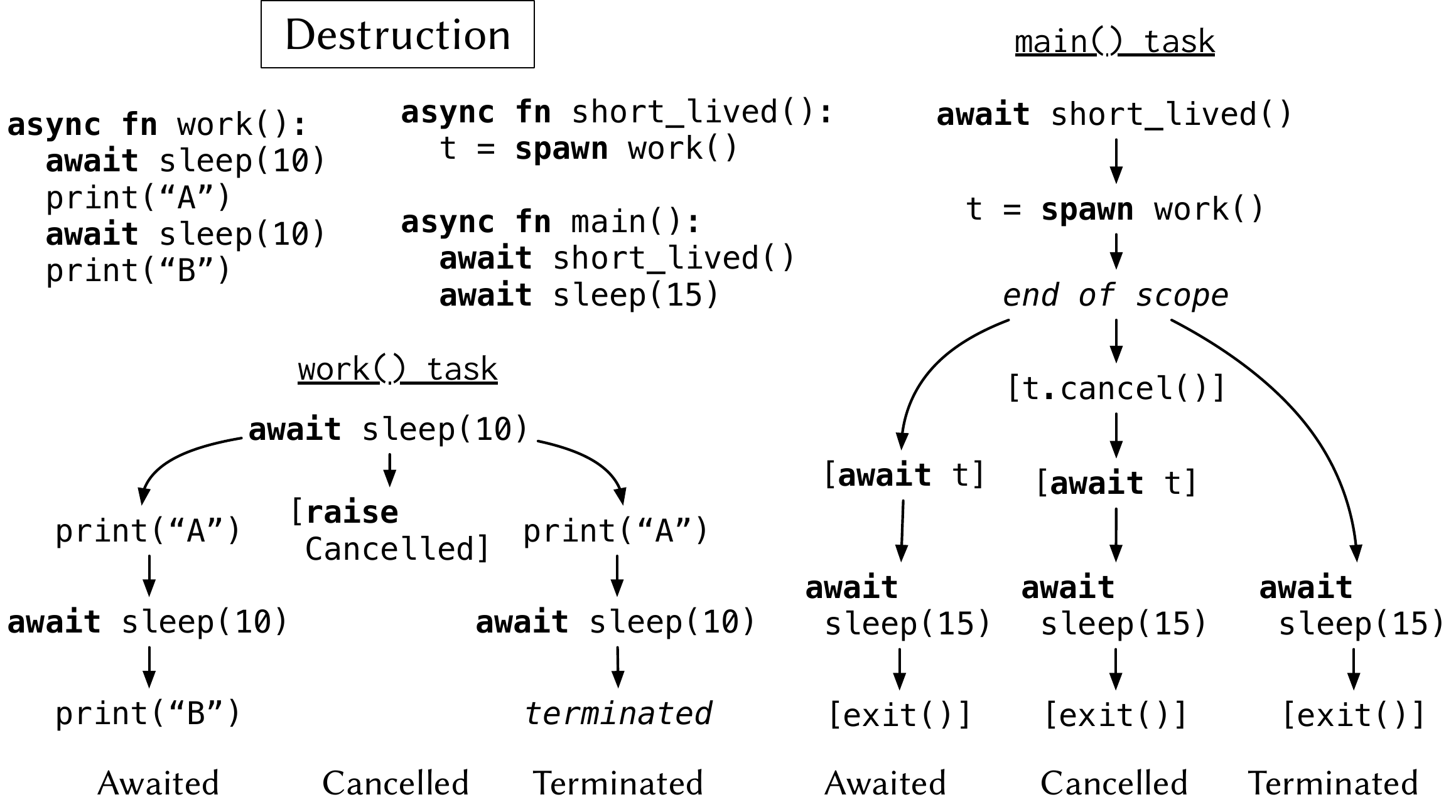}
  \caption{An example program demonstrating the \Destruction dimension. Each diagram shows the sequence of instruction execution for a task under Awaited (Python+Trio), Cancelled (Swift), and Terminated (\CSharp). The diagrams branch at the point of divergence between each approach. Administrative instructions that are executed but don't appear in the source code are displayed in brackets. }
  \label{fig:destruction}
\end{figure}

Once a language has decided that it is time to destroy a task, we encounter yet another point of divergence: what does it mean to destroy a task? For example, consider the program in \Cref{fig:destruction}. At the end of the \extent of the spawned \lstinline[language=Pseudo]|work()| task (either end of \lstinline[language=Pseudo]|main()| for \dyn \extent, or end of program for \indef \extent), what happens to the \lstinline[language=Pseudo]|work()| task? Today's languages offer three possibilities: to await the task, to cancel the task, or to terminate the task---the \awaited, \cancelled, and \terminated points of the \destruction dimension.

The \destruction dimension is orthogonal to \extent. \Js is a language with \indef \extent and \awaited \destruction{}---\Cref{fig:extent} previously showed an example relying on this pair of design decisions. Python+Trio is a language/system with \dyn \extent and \awaited \destruction, as tasks are tied to a ``nursery,'' and the nursery waits until all its child tasks are completed before exiting. The \awaited column in \Cref{fig:destruction} demonstrates Python+Trio semantics, where the program would print ``ABC'' by waiting for the \code|work()| task to finish before exiting \code|main()|.

Most other designs opt to cancel and then await\footnotemark{} tasks at the end of \extent, such as Swift, Rust+Tokio, Rust+Smol, and Python+Asyncio. \Cref{fig:destruction} running under Swift would output nothing because \code|work()| is cancelled before it can reach a print statement. By contrast, \CSharp will \emph{terminate} tasks at the end of their \indef \extent, meaning the runtime simply exits while the destroyed tasks are running. In \Cref{fig:destruction}, under \CSharp, the program would print ``A'' because the task continues until being terminated after 15 seconds.

\cprotect\footnotetext{The one exception here is Smol, which does not await cancelled tasks under its \dyn \extent by default. The reason is that Smol implements task destruction using Rust's \code|Drop| trait, which only permits synchronous code, so Rust async implementations cannot await the completion of a task on drop. Smol offers an async \code|Task::cancel| method that implements the cancel+await semantics, but this method is not called automatically. Rust+Tokio, by contrast, does implement cancel+await because it uses \indef \extent, and it is easier to have a runtime await outstanding tasks at the end of the main function. Libraries like Python+Trio can implement await-on-destruction using Python's \code|async with| construct for scoped async entry/exit methods.}

\paragraph{Design rationale.} For languages with \dyn \extent, the standard approach is to use \cancelled \destruction. The idea is that responsible async code should await on a task in the same scope in which it was spawned. If a task is still executing after its parent scope exits, this is probably unintended, and it could create surprising behavior to wait until that task is completed. However, this approach makes less sense in Python+Trio, whose nurseries do not expose any concept of an awaitable task handle, so it is not trivial to directly await the result of a spawned task.

For languages with \indef \extent, any approach can make sense depending on the needs of the application, and it is straightforward to configure a single global runtime to adopt the desired behavior. For example, Rust+Tokio can be configured to terminate rather than await.

\subsubsection{\Propagation}\label{ssec:designdims:propagation}

\begin{wrapfigure}{r}{0.37\linewidth}
\begin{lstlisting}[language=Py]
from trio import open_nursery

async def fail():
  raise Exception()

async def main():
  print("A")
  async with open_nursery() as n:
    n.start_soon(fail)
  print("B")
\end{lstlisting}
  \cprotect\caption{An example of \destruct \propagation. Python+Trio reraises exceptions when their associated nursery scope ends.}
  \label{fig:propagation:trio}
\end{wrapfigure}

One final point of divergence at a task's end of life is the handling of exceptions. In all surveyed languages with exceptions (Python, Swift, \Js, \CSharp), awaiting on an async function that has thrown an exception will re-raise the exception at the await point.\footnotemark{} However, what happens to an exception raised in a task that is never awaited? We call this the \propagation dimension.%
\cprotect\footnotetext{Instead of exceptions, Rust uses the \code|Result| type to communicate errors. An \code|Err| variant is returned on await, which we consider a semantically equivalent design to re-raising an exception at the await point.}

Most languages \never propagate, meaning the exception generates no behavior outside its containing task, apart from perhaps printing to the console. Python+Asyncio and \Js will print a stack trace for an uncaught asynchronous exception. Node.js\ specifically will exit with a non-zero error code, although this latter behavior is not part of the ECMAScript specification so we still categorize it as a \never approach to \propagation.

Python+Trio is the only system which offers a different semantics, in part because tasks cannot be awaited, so the library must offer another mechanism for observing asynchronous exceptions. The example in \Cref{fig:propagation:trio} (shown in Python, not pseudocode) will print ``A'' but not ``B'' because the nursery will re-raise the exception originating in the \code|fail| function. We describe this as \destruct \propagation. A similar program in any other system will print ``A'' then ``B'', possibly with a stack trace printed in-between.

\paragraph{Design rationale.} The upside to the \destruct approach is that an application is less likely to ignore problematic exceptions. With the \never approach, if a developer does not carefully monitor their logs, a task can fail and cause confusing bugs where expected operations never occur.

The downside is that the \destruct approach may run counter to either a developer's intuitions (i.e., exceptions aren't thrown away by the system) or their desired software architecture. Concurrency \abbr{api}s often seek to encapsulate errors at thread or task boundaries, so errors in one thread cannot crash the whole system. For example, in a backend web server with many concurrent connections, it is likely undesirable that one exceptional connection should crash the entire server. Either way, it seems sensible that a straight-line asynchrony \abbr{api} should offer \emph{some} way of accessing information about unawaited exceptions, rather than only printing to standard error, or ignoring unawaited exceptions outright.

\subsection{Cancellation}\label{ssec:designdims:cancellation}

\begin{wrapfigure}{r}{0.54\linewidth}
  \includegraphics[width=\linewidth]{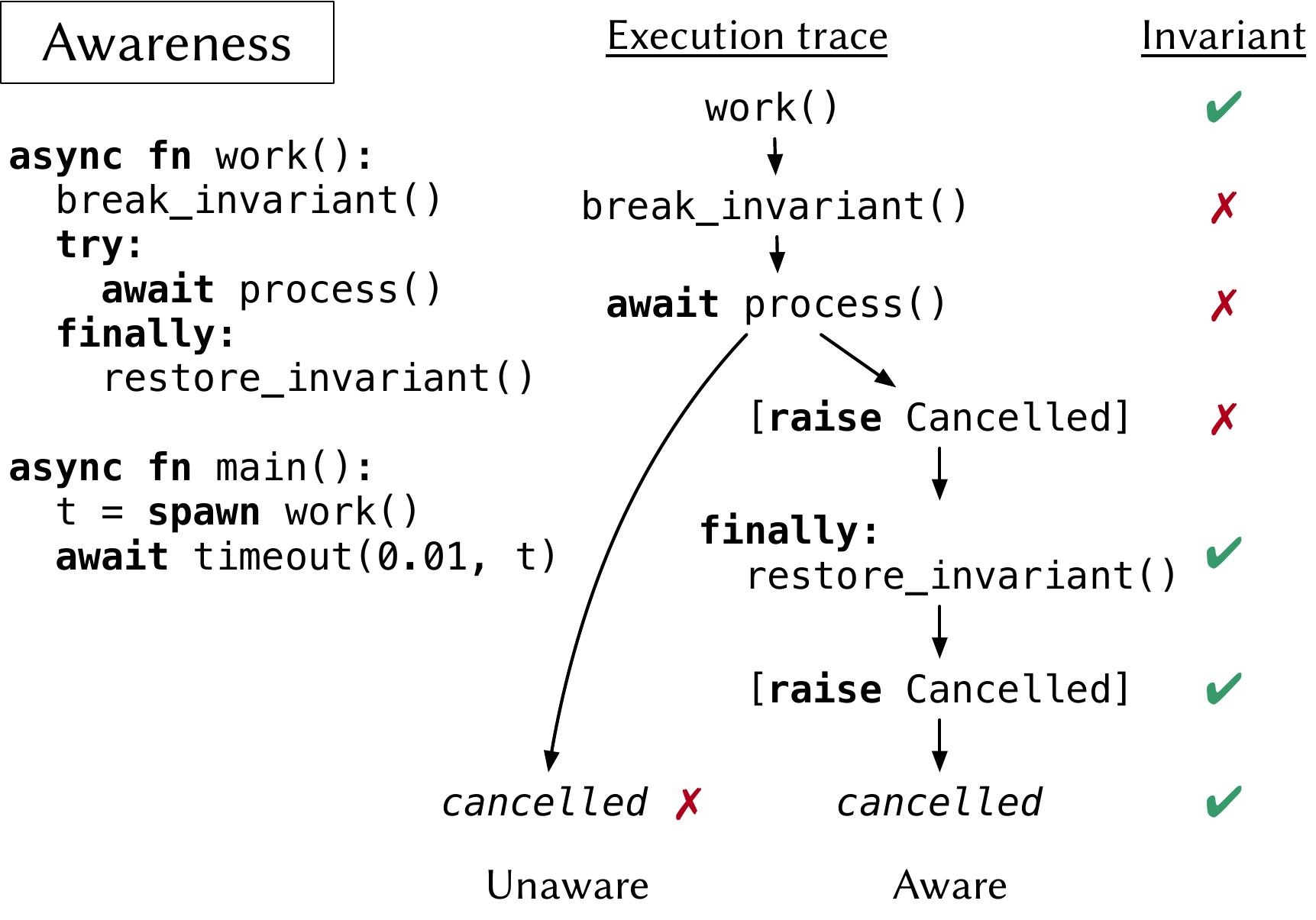}
  \caption{An example program demonstrating the \awareness dimension. A branching execution trace is paired with an indicator of whether an invariant is satisfied. An \aware approach allows the cancelled task to restore the invariant before proceeding with cancellation, while an \unaware approach remains cancelled with the invariant broken.}
  \label{fig:cooperation}
\end{wrapfigure}

As demonstrated by the case study on \code|timeout| (\Cref{sec:timeout}), cancellation is both a key component and a complicating factor in straight-line asynchrony. A naive cancellation approach would abruptly kill a running task, but no language adopts this approach because the task may be holding a mutex, writing to a file, or otherwise in a critical region with respect to a system's invariants.

A straightforward approach to cancellation in user-space is to use a \emph{cancellation token}, as demonstrated in \Cref{fig:timeout-tokio}. The developer is responsible for threading the token through all the relevant async code, and they are also responsible for reacting to the token in any async code that should be cancellable. Cancellation token \abbr{api}s can be found in \Js, \CSharp, and Rust. However, the amount of developer effort required for this approach is prohibitive. Modern takes on straight-line asynchrony have therefore sought to develop alternative cancellation approaches to lower the developer burden.

This subsection describes the two principal design dimensions of non-token-based, or \emph{integrated}, cancellation approaches, \awareness (whether a task can react to cancellation, \Cref{ssec:designdims:cooperation}) and \direction (how cancellation propagates through related tasks, \Cref{ssec:designdims:direction}). This subsection focuses on languages with integrated cancellation designs, namely Rust, Swift, and Python.

\subsubsection{\Awareness}
\label{ssec:designdims:cooperation}

Integrated cancellation approaches all center around triggering cancellations at await points. A key distinction is whether a cancelled task is aware that it is being cancelled, in order to handle or ignore the cancellation at a given await point.

The most straightforward approach is \unaware cancellation, where a task has no opportunity to react to being cancelled. This approach is taken by Rust in both Tokio and Smol. In Rust, cancellation essentially means never executing a coroutine again after it has yielded. With a direct handle on the coroutine, this simply means not polling the coroutine, or explicitly dropping it. If the coroutine is within a task, then the task must be marked as cancelled, and the runtime will similarly refuse to reschedule the task.

By contrast, Python and Swift both implement a form of \aware cancellation. In Python (both Asyncio and Trio), when a task is cancelled, a cancellation exception is thrown into the task. An async function can wrap any await point in a try/except block which can catch this cancellation exception, performing cleanup or consuming the exception.
In Swift, when a task is cancelled, a flag is set indicating that the task is cancelled. Functions can then opt in to raising a cancellation exception when this flag is set. It is encouraged to indicate a cancelled task by raising a \code|CancellationError|, which the standard library \code|Task.sleep| does.

\paragraph{Design rationale.} A benefit of \unaware cancellation is that tasks can be reliably cancelled. The principal case that delays \unaware cancellation is a compute-bound task which does not cooperatively yield to the runtime. With \aware cancellation, functions can either ignore cancellation or fail to opt in, which we discuss further in \Cref{ssec:designdims:direction}.

A major drawback to \unaware cancellation is that cancellation can easily violate logical invariants of a program. For example, consider the program in \Cref{fig:cooperation}. Say a program wants to do asynchronous work in a critical region where an invariant is broken and later restored. With \unaware cancellation, the \code|work| task can be halted at \lstinline[language=Pseudo]|await process()|, causing the invariant to remain broken. In Rust, this is a widely-documented problem with its cancellation system\footnotemark~\cite{paharia_400_2025,paharia_cancelling_2025}. With \aware cancellation through exceptions, the \lstinline[language=Pseudo]|finally| block has the opportunity to restore the invariant before continuing to raise the cancellation exception.

\cprotect\footnotetext{It is possible in Rust to implement something similar to the \code|finally| block using the \code|Drop| trait. When a task is cancelled, the task is eventually dropped and drop handlers are invoked for all coroutines in the task. One can theoretically define a custom \code|Drop| implementation for a bespoke type that calls \code|restore_invariant|. However, this approach is verbose (requiring a newtype and impl), unwieldy (requires mutable references to data structures, complicating ownership), and synchronous-only (one cannot call async code in a \code|Drop| implementation without blocking). }

\subsubsection{\Direction}
\label{ssec:designdims:direction}

Tasks form a dependency graph of dependents, tasks awaiting on others, and dependencies, tasks being awaited. Cancellation starts at a specific node, raising the question: how does cancellation proceed through the graph? We call approaches to this question the \direction dimension.

One approach is \tdown cancellation, where cancellation is communicated from dependents to dependencies. This approach describes Rust, which leans on the \code|Drop| trait to implement cascading communication during cancellation. In Smol, when a task is cancelled and later dropped, that drops the handles to all of that task's children, which in turn cancels those tasks, and so on through the task graph. This cancellation approach is feasible in Rust because it has a deterministic drop order, i.e., nodes have a well-defined parent / child relationship, where parents have an owning reference to the child.

Another approach is \bup cancellation, where cancellation is communicated from dependencies to dependents. This approach describes Python, which leans on exceptions to communicate cancellations. When a task is cancelled, both Asyncio and Trio will walk the task graph to find its leaf nodes, and raise a cancellation exception at those tasks. That exception is then raised across await points to async callers, which can choose to either catch the exception or allow it to propagate.

The final approach is \simultaneous cancellation, where cancellation is communicated at once to all transitive dependencies of the cancelled task. This approach describes Swift, which uses a shared cancellation flag. In Swift, convention holds that async functions should observe the cancellation flag and then raise a cancellation exception. A key difference from \bup cancellation is that the flag ensures that a task is aware of being cancelled, even if the cancellation exception was caught and consumed.

\paragraph{Design rationale.} The \tdown approach is a straightforward means of propagating cancellation without needing a mechanism like exceptions, which is especially useful for exception-less languages such as Rust. As previously mentioned, a \tdown approach requires the dependency graph to have a deterministic top-down order. The main limitation of the \tdown approach is that it makes custom handling of cancellations---such as running cleanup code---difficult: a cancelled task is never resumed, so it has no opportunity to observe its own cancellation, and cleanup is limited to synchronous destructors (e.g., Rust's \code{Drop} trait).

The \bup and \simultaneous approaches both provide the opportunity for handling cancellations. The major difference is that the \bup approach provides individual async functions more leeway to ignore cancellation, irrespective of the calling context. With the \simultaneous approach, every task in a task tree is guaranteed to know if it is cancelled. As implemented in Swift, this approach requires more complexity---cancellation is represented both formally through the cancellation flag, and conventionally through a cancellation exception observed by async primitives. In exchange, async functions can avoid the possibility of missed cancellations.


\subsubsection{\Persistence}
\label{ssec:designdims:persistence}

\begin{wrapfigure}{r}{0.62\linewidth}
  \includegraphics[width=\linewidth]{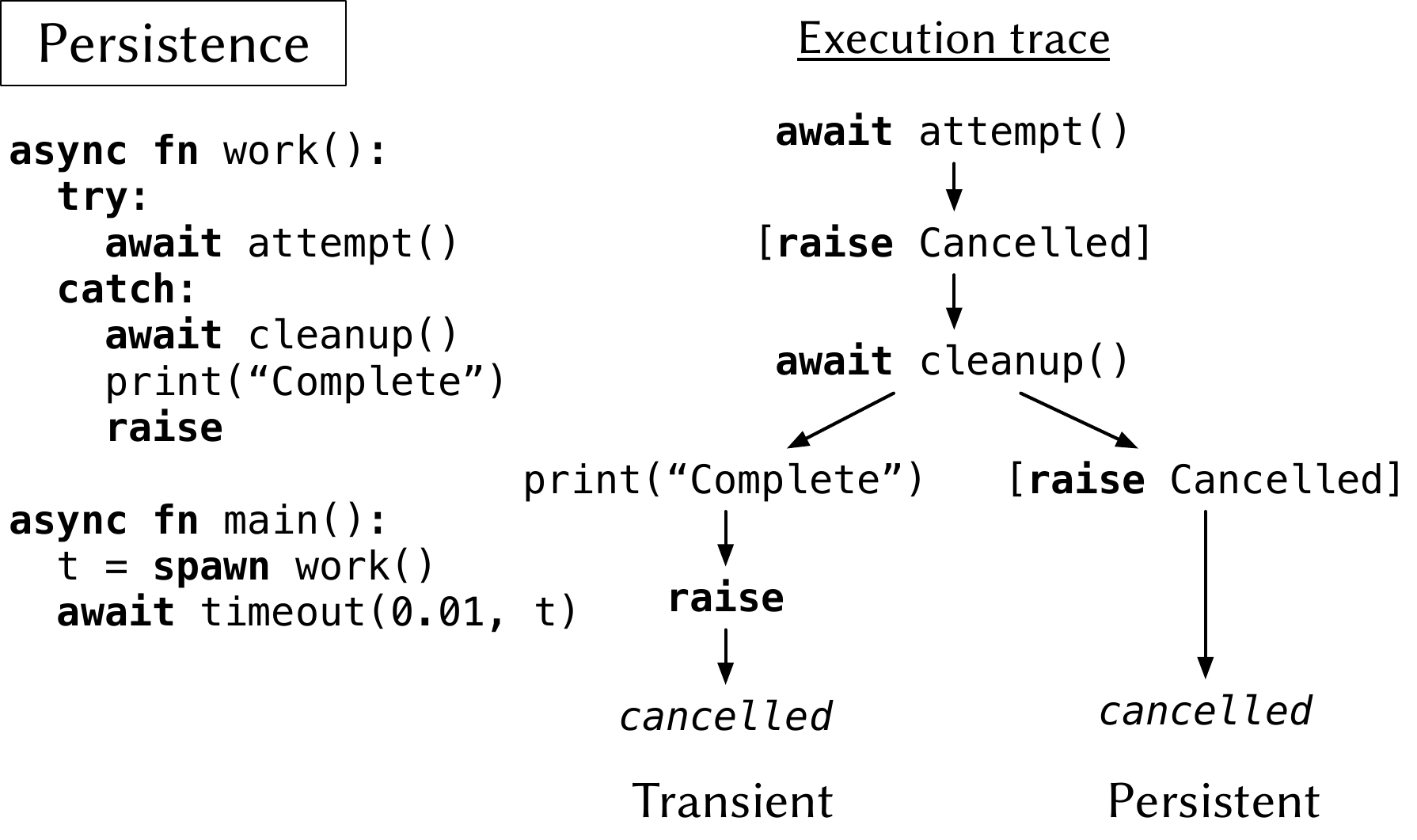}
  \cprotect\caption{An example program demonstrating \persistence. After the \code|work| task is cancelled, then the first await raises a cancellation exception. Under \Transient semantics (Python+Asyncio), subsequent awaits proceed as normal. Under \Persistent semantics (Python+Trio), subsequent awaits still raise a cancelled exception, stopping further execution.}
  \label{fig:persistence:shield}
\end{wrapfigure}

One final element of cancellation is its \persistence: when a task is cancelled, does the cancellation request persist (e.g., subsequent async work is also marked as cancelled), or is cancellation requested once? We will describe these options as \persistent cancellation and \transient cancellation. The \persistence dimension is important for cancel \aware languages, because on cancellation a \task may perform additional async work.

Swift and Python+Trio use \persistent cancellation. Swift and Trio both implement a cancellation flag shared amongst a task graph, and that flag is permanently active once set. Swift's flag is part of its public \abbr{api}, while Trio's flag is an implementation detail. Consequently, if an async computation ignores its cancellation, then subsequent awaits may again raise a cancellation exception; for both Swift and Trio any subsequent await point may result in a cancellation exception.

Rust and Python+Asyncio use \transient cancellation. In Rust's case, because cancellation is \noncoop, cancellation is unrecoverable. For Asyncio, cancellation is recoverable, so if an async computation ignores a cancellation request, then it can continue spawning and awaiting tasks unimpeded. Importantly, \tasks spawned after the initial cancellation request would be unaware of the previous cancellation attempt.

\paragraph{Design rationale.} For \noncoop cancellation, \persistence is a moot point, so the main question is which design best fits \coop cancellation. This perhaps comes down to a philosophy on why tasks would choose to ignore their cancellation, and what risks are incurred as a result.

The main argument for \transient cancellation is for \tasks to perform async resource cleanup. For example, as shown in \Cref{fig:persistence:shield}, if \code|work()| is cancelled in the middle of \code|attempt()|, then it can still call \code|await cleanup()| before re-raising the cancellation exception. A concrete instance of this pattern is that closing a \abbr{tls} socket may require asynchronously sending a \code|close_notify| alert to the peer so that they can be cryptographically assured that you intended to close the socket~\citeurlwith{https://docs.openssl.org/3.0/man3/SSL_shutdown/}{docs.openssl.org}.
However, \transient cancellation has a greater risk for cancellation signals to go unnoticed. If one \task suppresses a cancellation signal, then all others may never become aware of the attempt.%

Runtimes with \persistent cancellation typically come with a cancellation \emph{shielding} operator to allow for asynchronous cleanup. For example, the Python+Trio primitive \code|move_on_after| (demonstrated in \Cref{fig:timeout-simple-trio}) exposes a \code|shield| keyword argument that, when \code|True|, suppresses cancellation signals in nested code. This allows async cleanup code to temporarily ignore cancellation while still ensuring liveness by moving on after a fixed duration.

\section{Formal Model}\label{sec:semantics}

\begin{figure}[t]
  \begin{subfigure}[t]{0.4\textwidth}
    \vspace{0cm}
    \input{./f/lc-gr.tex}
    \cprotect\caption{A subset of the \lc grammar.
    \label{fig:lang:lc}}
  \end{subfigure}\hfill%
  \begin{minipage}[t]{0.55\textwidth}
    \vspace{0pt}
    \begin{subfigure}{\textwidth}
      \vspace{0cm}
      \input{./f/exn-gr.tex}
      \cprotect\subcaption{The extension grammar for \exn, which extends \lc with exception handling forms, including $\bsfid{throw-in}$, which we model after CPython's \code|.throw()| coroutine method.}
      \label{fig:lang:exn}
    \end{subfigure}
    \begin{subfigure}{\textwidth}
      \vspace{1em}
      \noindent
\begin{tabular*}{\linewidth}{@{} l @{\extracolsep{\fill}} r @{}}
  $\langle \sigma, E[\bsfid{throw-in}(\mathbf{fn}\ x \to E_{\mathit{inner}}[x], v)] \rangle$ &  [Throw-In] \\
$\longrightarrow \langle \sigma, E[E_{\mathit{inner}}[\bsfid{throw}\ v]] \rangle$ &
\end{tabular*}

      \cprotect\subcaption{An example reduction rule for the \exn extension language.}
      \label{fig:lang:exn:red}
    \end{subfigure}
  \end{minipage}
  \caption{An illustrative subset of the grammar and reduction rules for the core calculi \lc and \exn.}
  \label{fig:lang:core}
\end{figure}

The preceding taxonomy helps us make sense of the space of straight-line asynchrony, but it does not provide an account of each language's async holistically. In this section we model the core async features of each of the languages under study. An example of the payoff is \Cref{fig:explain-tree}, which precisely accounts for the divergent behaviors of \Cref{fig:motivating-example}. The rest of the section explains the key components of the figure.

Rather than describe each language separately, our semantics is designed to both capture the commonalities and highlight the differences between async in these languages. To do this, we unify the languages' implementation of asynchrony using delimited continuations rather than faithfully modeling each implementation using state machines and other such mechanisms. The full semantics is too large to include in its entirety; we provide the elided rules in the Supplemental Material.

We provide an executable version of the model in Redex~\cite{racket:redex}, including a full test suite that covers all the examples from \Cref{sec:motivation,sec:designdims} for all runtimes. The model includes a fuzzer that generates random programs, compiles them into the respective languages, and checks that the output from the real program and model program match. Due to non-determinism, we run each program $N=50$ times and check that the program outputs are a subset of model outputs. This gives us high confidence that our model matches the behavior of the corresponding runtimes.

\subsection{Base Language}

\Cref{fig:lang:lc,fig:lang:exn} provide the grammar and semantics of the core calculi that comprise our base for the async extensions. The core language \lc contains mutable references and delimited continuations via the $\bsfid{shift}$ and $\bsfid{reset}$ operators described by \citet{danvy_functional_nodate}. \Cref{fig:lang:lc} provides the grammar for \lc. Reduction rules for synchronous grammars are defined over a mutable heap, $\sigma$, and an expression $e$, $\langle\sigma,e\rangle$.

For languages with exceptions, we extend \lc with $\bsfid{try}$/$\bsfid{catch}$ and $\bsfid{throw}$ forms and refer to this language as \exn. \Cref{fig:lang:exn} provides the grammar extension for \exn, and \Cref{fig:lang:exn:red} provides the reduction rule for the nonstandard form $\bsfid{throw-in}$, which we model after Python's \code{.throw()} coroutine method. The form $\bsfid{throw-in}$ resumes a continuation with an exception instead of a value.

\begin{figure}[t]
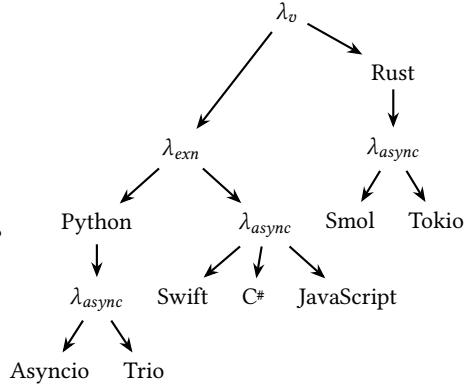

  \begin{subfigure}[t]{\textwidth}
    \input{./f/lasync-gr.tex}
    \cprotect\caption{An illustrative subset of the \lasync extension grammar.}
    \label{fig:lasync:grammar}
  \end{subfigure}\vspace{1em}
  \begin{subfigure}[t]{0.45\textwidth}
    \vspace{0cm}
    \input{./f/grammars.tex}
    \cprotect\caption{The grammar extensions for each language.}
    \label{fig:langs:exts}
  \end{subfigure}\hfill%
  \begin{subfigure}[t]{0.5\textwidth}
    \vspace{0cm}
    \begin{forest}
  for tree={
    edge={->, >={Stealth[length=2mm]}, thick, draw},
    l sep=0.3cm,
    s sep=0.2cm,
    font=\small,
  }
  [\lc
    [\exn, name=EXN
      [Python, name=Py
        [\leasync, name=EAsync_Py
          [Asyncio]
          [Trio]
        ]
      ]
      [\leasync, name=EAsync
        [Swift]
        [\CSharp]
        [\Js]
      ]
    ]
    [Rust, name=Rust, l-=1cm
      [\lasync, name=Async
        [Smol]
        [Tokio]
      ]
    ]
  ]
\end{forest}
    \cprotect\caption{The order of extension for each async runtime.}
    \label{fig:langs:lineage}
  \end{subfigure}
  \cprotect\caption{The grammar extensions and order of extension for \lasync and the seven async systems under study. Python and Rust are \emph{extended by} \lasync because they provide native coroutines. The runtimes for those languages are defined as libraries. The other languages: Swift, \Js, \CSharp, \emph{extend} \lasync because they do not provide native coroutines and their async runtimes are built into the language runtime.}
  \label{fig:platform}
\end{figure}

\subsection{The Async Platform}

We now describe the configuration of the shared async grammar, which can extend any \lc-based language. \Cref{fig:lasync:grammar} defines the shared grammar constructs that encode the async runtime building blocks discussed in \Cref{sec:designdims}. We will refer to this extension as the async platform, \lasync.

We introduce the notion of an async \emph{frame}, $F$, written $\langle \ell,\,e\rangle$, which consists of an expression $e$, along with a frame label, $\ell$.\footnotemark A \emph{frame label}, $\ell$, is either the name of the \task this frame corresponds to, or \emph{sync} to denote a synchronous frame. A \emph{frame stack}, $FS$, is a list of frames---essentially a thread. An empty frame stack is written $\epsilon$, and we write $F \circ FS$ to denote a frame stack whose head is $F$ and tail is $FS$. A \emph{process}, $P$, is a collection of frame stacks. We will use the syntax $P \uplus FS$ to select one frame stack in $P$ to run. A \emph{task queue}, $Q$, stores the async work waiting to be run; $Q$ comprises tuples that have a label and a continuation, $(\ell,\kappa)$.
A \emph{signals queue}, $T$, stores async work that is waiting on \abbr{i/o} operations. $T$ comprises tuples that have a time, a label, and a continuation, $(t,\ell,\kappa)$, $\kappa$, is invoked no earlier than timestamp $t$.
We will use the same stack notation for $Q$ and $T$ as we do for $FS$. We will additionally write $Q \qpush (\ell,\,\kappa)$ to append to the end of the queue.%
\footnotetext{We use a labeled async frame to explicitly associate executing code with a task. An alternative formalism would be to delimit tasks using a labeled context. We view these two options as equivalent and chose the former.}

The async platform provides a few additional administrative instructions that we will describe here. The $\bsfid{sys-block}(e)$ expression provides the barrier between synchronous code and asynchronous code. Languages may use a block expression to block a synchronous frame until the asynchronous expression has completed running. The block expression marks the runtime entry (e.g., creating a thread pool), and when it returns, the runtime exit (e.g., destroying the thread pool and runtime queues).

The expression $\bsfid{sys-io}(e_t,e)$ is the \abbr{i/o} operation provided by the runtime system. It models real-world \abbr{i/o} operations that would take some non-deterministic amount of time to resolve, like receiving packets from the network or querying a database. The form is prescient because the result is provided upfront: e.g., $\bsfid{sys-io}(5,\text{``hello''})$ creates a \task that will resolve to the value ``hello'' after at least five time steps. Furthermore, we model \abbr{i/o} operations that may not complete as $\bsfid{sys-io}(\infty, e)$. The $\bsfid{sys-io}$ form is deterministic in its evaluation \emph{value}, but not in its evaluation \emph{time}.

The expression $\bsfid{sys-start-soon}(\ell,\kappa)$ appends the continuation $\kappa$ to $Q$, set to run in a frame with the specified label $\ell$. The expression $\bsfid{sys-start-later}(\ell,\kappa)$ appends the continuation to the signal queue, $T$, which will be scheduled after timestep $t$.

The grammar extensions for the remaining languages are shown in \Cref{fig:langs:exts}, and the order in which we layer the extensions in \Cref{fig:langs:lineage}. The reduction rules for all async-capable languages, which base Python and Rust are not, are defined over the tuple $\langle t, \sigma, Q, T, P \rangle$. Most tuple items remain constant in many rules; to simplify the presentation we will use a double angle notation $\elided{\ldots}$ to indicate that unmodified tuple items are elided.

The goal of the operational semantics is to highlight the differences between languages. We will therefore present the same reduction rule from different languages together. We will also call attention to the portion relevant to each design dimension with a highlight in the following colors: \ceag{\eagerness}, \cext{\extent}, \cdes{\destruction}, \cpro{\propagation}, and \csus{\suspension}, and within cancellation: \cawa{\awareness}, \cdir{\direction}, and \cper{\persistence}. For space reasons we do not present the rules relevant for \references, and we simplify the presentation with a liberal use of metafunctions and macros, see \Cref{table:metafunctions} for a description of those used; the full definitions of grammars, reductions, and metafunctions are provided in the Supplemental Material.

\begin{table}[t]
  \small
\begin{tabularx}{\linewidth}{@{}l>{\raggedright\arraybackslash}X@{}}
\toprule
Metafunction & Description \\
\midrule
$\sfid{task-cancelled} : \sigma \times \ell \to b$ & $\mathit{true}$ iff task $\ell$, or any transitive dependent, is cancelled \\
$\sfid{task-alloc} : \sigma \to (\sigma, x)$ & Allocate a new task struct bound under gensymed name $x$ in $\sigma$ \\
$\sfid{task-uncancel} : \sigma \times \ell \to \sigma$ & Remove cancellation from task $\ell$ in $\sigma$ \\
$\sfid{task-add-dependency} : \sigma \times \ell_1 \times \ell_2 \to \sigma$ & Add task $\ell_2$ as a dependency of $\ell_1$ \\
\midrule
Macro & Description \\
\midrule
$\semmacro{task-add-dependent} : v \times \ell \times e \rightsquigarrow e$ & Register $\ell$ (continuation $e$) as a dependent of awaitable $v$ \\
$\semmacro{task-set-done} : x \times e \rightsquigarrow e$ & Evaluate $e$ and use value as $x$'s resolved success value \\
$\semmacro{task-set-failed} : x \times e \rightsquigarrow e$ & Evaluate $e$ and use value as $x$'s resolved failure value \\
$\semmacro{task-is-completed} : e \rightsquigarrow e$ & $\mathit{true}$ iff task is resolved (success or failure) \\
$\semmacro{task-get-result} : e \rightsquigarrow e$ & Return task's final value; reraise if task failed \\
$\semmacro{task-continue-with} : v \times x \rightsquigarrow e$ & Resume continuation bound by $x$ with settled value of task $v$ \\
$\semmacro{task-get-dependents} : e \rightsquigarrow e$ & Return dependents of the task that $e$ evaluates to \\
$\semmacro{task-cancel-dependencies} : x \rightsquigarrow e$ & Cancel task $x$ and its descendants \\
$\semmacro{task-wait-on-dependencies} : x \rightsquigarrow e$ & Wait for all dependencies of $x$ to settle \\
$\semmacro{task-reraise-dependency-failures} : x \rightsquigarrow e$ & Reraise dependency failures \\
\bottomrule
\end{tabularx}

  \caption{Metafunctions and macros used in the reduction rules. Metafunctions evaluate to a term, while macros expand into expressions. We set macros in italics to distinguish them from metafunctions.}
  \label{table:metafunctions}
  \vspace{-2em}
\end{table}

\subsection{Start/End of Life}

\begin{figure}[t]
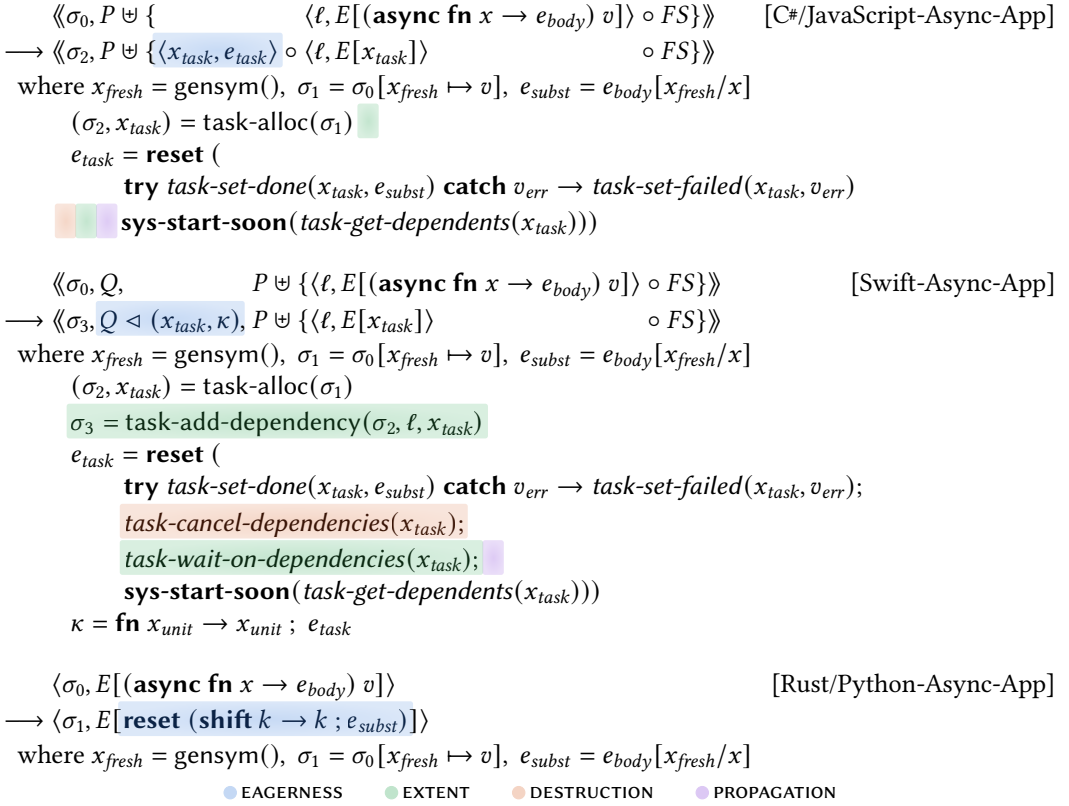

    \noindent
$\begin{array}[t]{@{}r@{\ }l@{}l@{}l@{}l@{}}
                & \elided{\sigma_0, P \uplus \{ & & \langle \ell, E[(\bsfid{async\ fn}\ \many{x} \to e_{\mathit{body}})\ \many{v}] \rangle & {}\circ FS \}} \\
\longrightarrow & \elided{\sigma_2, P \uplus \{ \ceag{\langle x_{\mathit{task}}, e_{\mathit{task}} \rangle} & {}\circ{} & \langle \ell, E[x_{\mathit{task}}] \rangle & {}\circ FS \}}
\end{array}$\hfill$\text{[\CSharp/\Js-Async-App]}$\par
\begin{tabular*}{\linewidth}{@{} l @{\extracolsep{\fill}} r @{}}
\multicolumn{2}{@{}l@{}}{
  $\begin{array}{l}
  \where\ \many{x}_{\mathit{fresh}} = \many{\sfid{gensym}(),~ \sigma_1 = \sigma_0[\many{x}_{\mathit{fresh}} \mapsto \many{v}],~ e_{\mathit{subst}} = e_{\mathit{body}}[\many{x}_{\mathit{fresh}}/\many{x}]} \\
  \qquad (\sigma_2, x_{\mathit{task}}) = \sfid{task-alloc}(\sigma_1)\;\cext{\ghost} \\
  \qquad e_{\mathit{task}} = \bsfid{reset}\ ( \\
  \qquad \qquad \bsfid{try}\ \semmacro{task-set-done}(x_{\mathit{task}}, e_{\mathit{subst}})\ \bsfid{catch}\ v_{\mathit{err}} \to \semmacro{task-set-failed}(x_{\mathit{task}}, v_{\mathit{err}}) \\
  \quad\;\;\cdes{\ghost}\;\cext{\ghost}\;\cpro{\ghost}\;
  \bsfid{sys-start-soon}(\semmacro{task-get-dependents}(x_{\mathit{task}})))
  \end{array}$
}
\end{tabular*}

\vspace{1em}
\noindent
$\begin{array}[t]{@{}r@{\ }l@{\ }l@{}l@{}}
                & \elided{\sigma_0, Q, & P \uplus \{ \langle \ell, E[(\bsfid{async\ fn}\ \many{x} \to e_{\mathit{body}})\ \many{v}] \rangle & {}\circ FS \}} \\
\longrightarrow & \elided{\sigma_3, \ceag{Q \qpush (x_{\mathit{task}}, \kappa)}, & P \uplus \{ \langle \ell, E[x_{\mathit{task}}] \rangle & {}\circ FS \}}
\end{array}$\hfill$\text{[Swift-Async-App]}$\par
\begin{tabular*}{\linewidth}{@{} l @{\extracolsep{\fill}} r @{}}
\multicolumn{2}{@{}l@{}}{
  $\begin{array}{l}
  \where\ \many{x}_{\mathit{fresh}} = \many{\sfid{gensym}(),~ \sigma_1 = \sigma_0[\many{x}_{\mathit{fresh}} \mapsto \many{v}],~ e_{\mathit{subst}} = e_{\mathit{body}}[\many{x}_{\mathit{fresh}}/\many{x}]} \\
  \qquad (\sigma_2, x_{\mathit{task}}) = \sfid{task-alloc}(\sigma_1) \\
  \qquad \cext{\sigma_3 = \sfid{task-add-dependency}(\sigma_2, \ell, x_{\mathit{task}})} \\
  \qquad e_{\mathit{task}} = \bsfid{reset}\ ( \\
  \qquad \qquad \bsfid{try}\ \semmacro{task-set-done}(x_{\mathit{task}}, e_{\mathit{subst}})\ \bsfid{catch}\ v_{\mathit{err}} \to \semmacro{task-set-failed}(x_{\mathit{task}}, v_{\mathit{err}}) \mathbin{;} \\
  \qquad \qquad \cdes{\semmacro{task-cancel-dependencies}(x_{\mathit{task}}) \mathbin{;}} \\
  \qquad \qquad \cext{\semmacro{task-wait-on-dependencies}(x_{\mathit{task}}) \mathbin{;}}\;\cpro{\ghost}\\
  \qquad \qquad \bsfid{sys-start-soon}(\semmacro{task-get-dependents}(x_{\mathit{task}}))) \\
  \qquad \kappa = \bsfid{fn}\ x_\mathit{unit} \to x_\mathit{unit}\mathbin{;}\ e_{\mathit{task}}
  \end{array}$
}
\end{tabular*}

\vspace{1em}
\noindent
$\begin{array}[t]{@{}r@{\ }l@{}}
                & \langle \sigma_0, E[(\bsfid{async\ fn}\ \many{x} \to e_{\mathit{body}})\ \many{v}] \rangle \\
\longrightarrow & \langle \sigma_1, E[\ceag{\bsfid{reset}\ (\bsfid{shift}\ k \to k \mathbin{;} e_{\mathit{subst}})}] \rangle
\end{array}$\hfill$\text{[Rust/Python-Async-App]}$\par
\begin{tabular*}{\linewidth}{@{} l @{\extracolsep{\fill}} r @{}}
\multicolumn{2}{@{}l@{}}{
  $\begin{array}{l}
  \where\ \many{x}_{\mathit{fresh}} = \many{\sfid{gensym}(),~ \sigma_1 = \sigma_0[\many{x}_{\mathit{fresh}} \mapsto \many{v}],~ e_{\mathit{subst}} = e_{\mathit{body}}[\many{x}_{\mathit{fresh}}/\many{x}]}
  \end{array}$
}
\end{tabular*}

    \vspace{-0.5em}
    \makelegend{eag,ext,des,pro}
    \cprotect\caption{A comparison of the rules \semrule{Async-App} across languages.}
    \label{fig:sem:app}
\end{figure}

\begin{figure}[tb]
    \noindent
$\begin{array}[t]{@{}r@{\ }l@{\ }l@{}l@{}}
                & \elided{ \sigma_0, Q, & P \uplus \{ \langle \ell, E[\bsfid{spawn}\ v_{\mathit{coro}}] \rangle & {}\circ FS \}} \\
\longrightarrow & \elided{ \sigma_1, \ceag{Q \qpush (x_{\mathit{task}}, \kappa)}, & P \uplus \{ \langle \ell, E[x_{\mathit{task}}] \rangle & {}\circ FS \}}
\end{array}$\hfill$\text{[Asyncio/Tokio/Smol-Spawn]}$\par
\begin{tabular*}{\linewidth}{@{} l @{\extracolsep{\fill}} r @{}}
\multicolumn{2}{@{}l@{}}{
  $\begin{array}{l}
  \where\ (\sigma_1, x_{\mathit{task}}) = \sfid{task-alloc}(\sigma_0) \\
  \qquad e_{\mathit{task}} = \bsfid{reset}\ ( \\
  \qquad \qquad \difff{\bsfid{try}}\ \semmacro{task-set-done}(x_{\mathit{task}}, \bsfid{await}\ v_{\mathit{coro}})\ \difff{\bsfid{catch}\ v_{\mathit{err}} \to \semmacro{task-set-failed}(x_{\mathit{task}}, v_{\mathit{err}})} \mathbin{;} \\
  \quad\;\;\cdes{\ghost}\;\cext{\ghost}\;\cpro{\ghost}\;
  \bsfid{sys-start-soon}(\semmacro{task-get-dependents}(x_{\mathit{task}}))) \\
  \qquad \kappa = \bsfid{fn}\ x_{\mathit{unit}} \to x_{\mathit{unit}} \mathbin{;} e_{\mathit{task}}
  \end{array}$
}
\end{tabular*}

\vspace{1em}
\noindent
$\begin{array}[t]{@{}r@{\ }l@{\ }l@{}l@{}}
                & \elided{ \sigma_0, Q, & P \uplus \{ \langle \ell, E[\bsfid{spawn}\ v_{\mathit{coro}}] \rangle & {}\circ FS \}} \\
\longrightarrow & \elided{ \sigma_2, \ceag{Q \qpush (x_{\mathit{task}}, \kappa)}, & P \uplus \{ \langle \ell, E[x_{\mathit{task}}] \rangle & {}\circ FS \}}
\end{array}$\hfill$\text{[Trio-Spawn]}$\par
\begin{tabular*}{\linewidth}{@{} l @{\extracolsep{\fill}} r @{}}
\multicolumn{2}{@{}l@{}}{
  $\begin{array}{l}
  \where\ (\sigma_1, x_{\mathit{task}}) = \sfid{task-alloc}(\sigma_0) \\
    \qquad \cext{\sigma_2 = \sfid{task-add-dependency}(\sigma_1, \ell, x_{\mathit{task}})} \\
  \qquad e_{\mathit{task}} = \bsfid{reset}\ ( \\
  \qquad \qquad \bsfid{try}\ \semmacro{task-set-done}(x_{\mathit{task}}, \bsfid{await}\ v_{\mathit{coro}})\ \bsfid{catch}\ v_{\mathit{err}} \to \semmacro{task-set-failed}(x_{\mathit{task}}, v_{\mathit{err}}) \mathbin{;} \\
  \qquad \quad\;\cdes{\ghost}\;\cext{\semmacro{task-wait-on-dependencies}(x_{\mathit{task}}) \mathbin{;}}\\
  \qquad \qquad \cpro{\semmacro{task-reraise-dependency-failures}(x_{\mathit{task}}) \mathbin{;}} \\
  \qquad \qquad \bsfid{sys-start-soon}(\semmacro{task-get-dependents}(x_{\mathit{task}}))) \\
  \qquad \kappa = \bsfid{fn}\ x_{\mathit{unit}} \to x_{\mathit{unit}} \mathbin{;} e_{\mathit{task}}
  \end{array}$
}
\end{tabular*}

    \makelegend{eag,ext,des,pro}
    \cprotect\caption{A comparison of \semrule{Spawn} across Rust/Python runtimes. The $\bsfid{try/catch}$ is colored gray because it is present in the Asyncio semantics, but not those for Tokio/Smol---the rules are otherwise equivalent.}
    \label{fig:sem:spawn}
\end{figure}

The rule \semrule{Async-App} defines how a \coro/\task is constructed and, for non-\lazy languages, what a \task does at the end of its computation. We will use the definitions in \Cref{fig:sem:app} to highlight the semantics for the start and end of a computation's life.

\Cref{fig:sem:app} provides the \semrule{Async-App} rule for \eager, \seager, and \lazy async applications. For \eager and \seager languages an async application follows a similar procedure: a new \task, $x_{\mathit{task}}$, is allocated in the global store; evaluation of the function body is wrapped in a try/catch block, on successful evaluation the result is stored in the task and it's marked as completed, on exception the error value is stored in the task and it's marked as failed. After the task is marked as completed or failed, all dependent tasks, those awaiting on the current \task, are scheduled to resume executing with the $\bsfid{sys-start-soon}$ form. \Eager and \seager languages differ by where the new $e_{\mathit{task}}$ expression is placed. \Eager languages set the expression in a new frame on \emph{top} of the current frame stack. \Seager languages place the expression in the task queue, $Q$, to run soon by the next available thread. The semantics for the \lazy languages diverge significantly. Neither Rust nor Python has a built-in async runtime, so their semantics are not defined as an extension of \lasync. The \lazy \semrule{Async-App} variant simply captures the function body as a continuation. We will discuss the \semrule{Spawn} rule---used to turn a \coro into a \task{}---below.

\semrule{Async-App} for the (semi-)eager languages also defines the end of life for a \task, including the \extent, \destruction, and \propagation semantics. When \CSharp/\Js allocate a \task, it is independent of the current context, meaning that the new \task has no relationship to the currently executing \task (labelled $\ell$). The absence of a relationship between the two \tasks marks the \indef \extent semantics for these two languages. In contrast, allocated \tasks in Swift are added as dependencies of the currently executing \task. Furthermore, within a Swift task body, $e_{\mathit{task}}$, after the \task is completed or failed its dependencies are cancelled and waited on. The dependency between the two tasks and the added operations at the end of the \task body marks the different semantics in Swift along the \extent and \destruction dimensions. Although Swift cancels and waits for dependencies to finish, exceptions held by \emph{failed} \tasks are not propagated.

\begin{figure}[tb]
  \noindent
$\begin{array}[t]{@{}r@{\ }l@{}l@{}}
                & \elided{ P \uplus \{ \langle \ell, E[\bsfid{await}\ v_{\mathit{aw}}] \rangle & {}\circ FS \}} \\
\longrightarrow & \elided{ P \uplus \{ \langle \ell, E[e_{\mathit{await}}] \rangle & {}\circ FS \}}
\end{array}$\hfill$\text{[\CSharp/Swift/Asyncio/Trio/Smol-Await]}$\par
\begin{tabular*}{\linewidth}{@{} l @{\extracolsep{\fill}} r @{}}
\multicolumn{2}{@{}l@{}}{
  $\begin{array}{l}
  \where\ e_{\mathit{await}} = \begin{array}[t]{@{}l@{\;}l@{}}
      \bsfid{if}   & \semmacro{task-is-completed}(v_{\mathit{aw}})\ \bsfid{then}\ \semmacro{task-get-result}(v_{\mathit{aw}}) \\
      \bsfid{else} & {\csus{\bsfid{shift}\ k \to}}\ \semmacro{task-add-dependent}(v_{\mathit{aw}},
\ell, \semmacro{task-continue-with}(v_{\mathit{aw}}, k))
    \end{array}
  \end{array}$
}
\end{tabular*}

\vspace{1em}
\noindent
$\begin{array}[t]{@{}r@{\ }l@{}l@{}}
                & \elided{ P \uplus \{ \langle \ell, E[\bsfid{await}\ v_{\mathit{aw}}] \rangle & {}\circ FS \}} \\
\longrightarrow & \elided{ P \uplus \{ \langle \ell, E[e_{\mathit{await}}] \rangle & {}\circ FS \}}
\end{array}$\hfill$\text{[\Js-Await]}$\par
\begin{tabular*}{\linewidth}{@{} l @{\extracolsep{\fill}} r @{}}
\multicolumn{2}{@{}l@{}}{
  $\begin{array}{l}
  \where\ e_{\mathit{await}} = \begin{array}[t]{@{}l@{\;}l@{\;}l@{}}
    {\csus{\bsfid{shift}\ k
    \to}} & \bsfid{if}   & \semmacro{task-is-completed}(v_{\mathit{aw}})\ \bsfid{then}\ \semmacro{task-get-result}(v_{\mathit{aw}}) \\
           & \bsfid{else} & \semmacro{task-add-dependent}(v_{\mathit{aw}},
\ell, \semmacro{task-continue-with}(v_{\mathit{aw}}, k))
    \end{array}
  \end{array}$
}
\end{tabular*}

  \makelegend{sus}
  \cprotect\caption{A comparison of \semrule{Await} semantics that define \dynamic and \static \suspension. \Dynamic suspending languages only suspend when the awaited value isn't immediately ready.}
  \label{fig:sem:await}
\end{figure}

The rule \semrule{Spawn} is provided by the runtimes of \lazy languages to elevate a \coro to a \task registered in the runtime. \semrule{Spawn} defines the semantics for the \task start and end of life that are absent from the \lazy definition of \semrule{Async-App}.

\semrule{Spawn} is defined separately for languages with \indef and \dyn \extent. The definition used by Asyncio/Tokio/Smol allows for \indef \extent. The rule follows the semantics discussed previously for \CSharp's \semrule{Async-App} rule. Trio enforces \dyn \extent and the rule follows closely that discussed previously for Swift. However, after the task body, $e_{\mathit{task}}$, finishes executing, dependencies are waited on \emph{uncancelled} to finish, and exceptions from failed dependencies are reraised. These differences mark Trio's unique combination of \awaited \destruction and \destruct \propagation semantics.

\Cref{fig:sem:await} provides the definitions for the rule \semrule{Await}. \Js, the only language sampled with \static \suspension semantics, always captures the current delimited continuation and adds itself as a dependent on the awaited value. The captured continuation, $e_{\mathit{k}}$, is returned. Languages with \dyn \suspension first check if the awaited value is ready---if it is, no suspension occurs. If a value is not ready, the continuation is captured and registered as a dependent of the pending computation; $e_{\mathit{k}}$ is returned.

\subsection{Cancellation}

\begin{figure}
  \noindent
$\begin{array}[t]{@{}r@{\ }l@{}l@{}}
                & \elided{ P \uplus \{ \langle \ell, E[\cdir{\bsfid{cancelled?}()}] \rangle & {}\circ FS \}} \\
\longrightarrow & \elided{ P \uplus \{ \langle \ell, E[\cawa{e_{\mathit{bool}}}] \rangle & {}\circ FS \}}
\end{array}$\hfill$\text{[Swift-Cancelled?]}$\par
\begin{tabular*}{\linewidth}{@{} l @{\extracolsep{\fill}} r @{}}
\multicolumn{2}{@{}l@{}}{
  $\begin{array}{l}
    \where\ e_{\mathit{bool}} = \sfid{task-cancelled}(\sigma, \ell)\,\cper{\ghost}
  \end{array}$
}
\end{tabular*}

\vspace{1em}
\noindent
$\begin{array}[t]{@{}r@{\ }l@{}l@{}l@{\ }l@{}l@{}l@{}}
                & \elided{(\ell, \kappa) & {}\circ{} & Q, & P \uplus \{ & & \epsilon \}}\\
\longrightarrow & \elided{               &           & Q, & P \uplus \{ \cawa{\langle \ell,}\kern3.5pt \cdir{\bsfid{throw-in}(\kappa, \textsf{``stop''})}\kern3.5pt \cawa{\rangle} & {}\circ{} & \epsilon \}}
\end{array}$\hfill$\text{[Trio-Schedule-Cancelled]}$\par
\begin{tabular*}{\linewidth}{@{} l @{\extracolsep{\fill}} r @{}}
\multicolumn{2}{@{}l@{}}{
  $\begin{array}{l}
    \where\ \bsfid{true} = \sfid{task-cancelled}(\sigma, \ell)\,\cper{\ghost}
  \end{array}$%
}
\end{tabular*}

\vspace{1em}
\noindent
$\begin{array}[t]{@{}r@{\ }l@{}l@{}l@{\ }l@{}l@{}l@{}}
                & \elided{\sigma_0, (\ell, \kappa) & {}\circ{} & Q, & P \uplus \{ & & \epsilon \}} \\
  \longrightarrow & \elided{\sigma_1,                &           & Q, & P \uplus \{ \cawa{\langle \ell,}\kern3.5pt \cdir{\bsfid{throw-in}(\kappa, \textsf{``stop''})}\kern3.5pt \cawa{\rangle} & {}\circ{} & \epsilon \}}
\end{array}$\hfill$\text{[Asyncio-Schedule-Cancelled]}$\par
\begin{tabular*}{\linewidth}{@{} l @{\extracolsep{\fill}} r @{}}
\multicolumn{2}{@{}l@{}}{
  $\begin{array}{l}
  \where\ \bsfid{true} = \sfid{task-cancelled}(\sigma_0, \ell),~
  \cper{\sigma_1 = \sfid{task-uncancel}(\sigma_0, \ell)}
  \end{array}$
}
\end{tabular*}

\vspace{1em}
\noindent
$\begin{array}[t]{@{}r@{\ }l@{}l@{}l@{}l@{}l@{}}
                & \elided{\sigma, (\ell, \kappa) & {}\circ{} & Q, P \uplus \{ & & \epsilon \}} \\
\longrightarrow & \elided{\sigma,                &           & Q, P \uplus \{ \langle \ell, \cawa{e_{\mathit{fail}}} \rangle & {}\circ{} & \epsilon \}}
\end{array}$\hfill$\text{[Tokio/Smol-Schedule-Cancelled]}$\par
\begin{tabular*}{\linewidth}{@{} l @{\extracolsep{\fill}} r @{}}
\multicolumn{2}{@{}l@{}}{
  $\begin{array}{l}
  \where\ \sfid{true} = \sfid{task-cancelled}(\sigma, \ell)\, \cper{\ghost} \\
  \qquad e_{\mathit{fail}} = \semmacro{task-set-failed}(\ell, ()) \mathbin{;} \bsfid{sys-start-soon}(\semmacro{task-get-dependents}(\ell))
  \end{array}$
}
\end{tabular*}

  \makelegend{awa,dir,per}
  \cprotect\caption{An illustrative subset of cancellation semantics.}
  \label{fig:sem:cancel}
\end{figure}

\Cref{fig:sem:cancel} provides the full set of rules for the cancellation mechanism provided by Swift, Tokio, Smol, Asyncio, and Trio. No rules are present for \CSharp or \Js as they do not provide a built-in mechanism to cancel a running \task (see \Cref{ssec:designdims:cancellation}). The design dimensions for cancellation are \awareness, \direction, and \persistence.

The \semrule{Cancelled?} rule is provided for Swift to allow a running \task to check if it has been cancelled. This form defines both the \aware \awareness and \simultaneous \direction that is unique to Swift. A \task does not need to rely on another mechanism to signal cancellation; this information is communicated between the runtime directly to each \task.

The \semrule{Schedule-Cancelled} rules for Asyncio/Trio are almost identical. When a cancelled \task is rescheduled the runtime will resume the continuation with an exception, using our $\bsfid{throw-in}$ form. By resuming the computation they are \aware of their cancellation. The cancellation \direction flows \bup because the \tasks in the task queue, $Q$, are depended on by others and not awaiting for any dependencies; in other words, they are the deepest dependencies.


Where Asyncio and Trio diverge is cancellation \persistence. A peculiarity of Asyncio is the \transient \persistence of cancellation: after a cancelled \task is resumed with a cancellation signal, it is no longer marked as cancelled. Further async work can be done and a second cancellation signal will not be raised. Trio and Swift both have \persistent cancellation, so further async work in an exception handler would be resumed with another cancellation exception.\footnote{To perform asynchronous cleanup work, runtimes like Trio and Swift will provide cancellation \emph{shields}, which temporarily protect a \task against cancellation so that async work can finish.}

The Rust runtimes, Tokio and Smol, do not provide cancellation \awareness. Cancelled \tasks are not rescheduled but deallocated: the continuation $\kappa$ is thrown away and the \task is set as failed by the expression $e_{\mathit{fail}}$.

\subsection{Seven Runtimes, One Program}

\begin{figure}[htb]
  \input{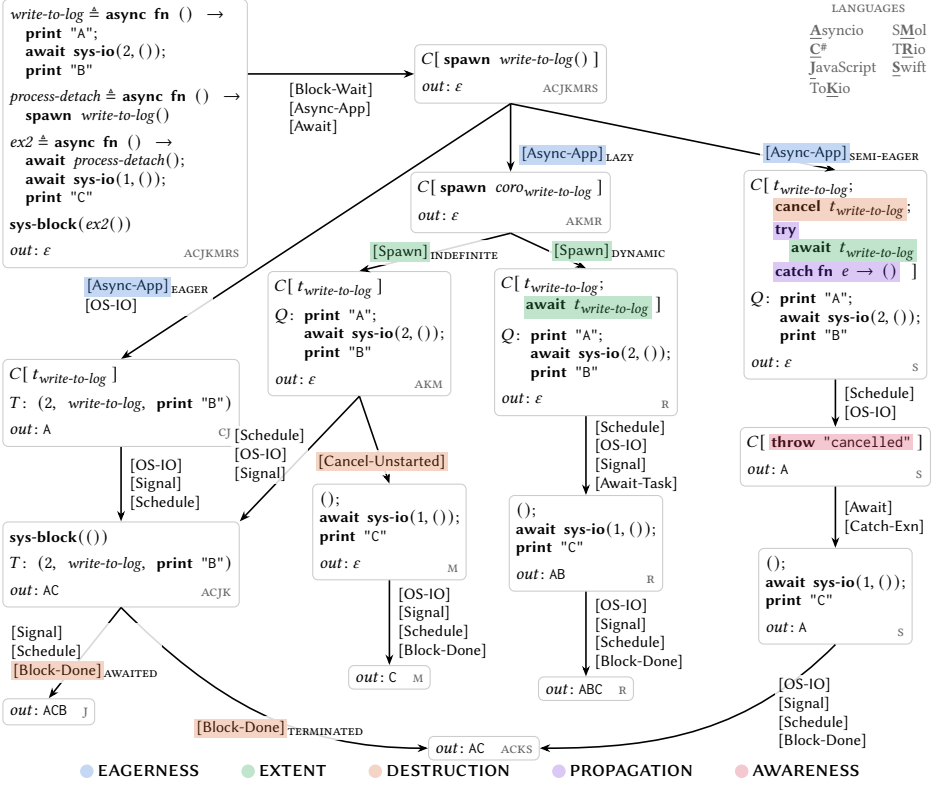}
  \makelegend{eag,ext,des,pro,awa}
  \cprotect\caption{The branching evaluation of program \code|ex2| from \Cref{fig:motivating-example}. The nodes in the \abbr{dag} represent the program state at that point. Between nodes we perform multiple reductions until the semantics of the runtimes diverge. A node has several outgoing edges when the runtimes disagree on the next applicable rule and we highlight the rules with the color of the design dimension along which they diverge; we indicate which design point each branch uses with a subscript. For example, on the rightmost path, Swift's \seager \semrule{Async-App} queues $\iid{write-to-log}$ and injects a scope-exit epilogue that cancels its dependency (\cdes{\destruction}), awaits it (\cext{\extent}), and discards any resulting exception (\cpro{\propagation}). When $\iid{write-to-log}$ runs, its cancellation-aware \semrule{OS-IO} implementation observes the flag and raises a \emph{cancelled} exception (\cawa{\awareness}).}
  \label{fig:explain-tree}
\end{figure}

\Cref{fig:explain-tree} brings the design dimensions and the formal model full circle: it evaluates program \code|ex2| from \Cref{fig:motivating-example} under all seven runtimes at once, branching precisely where their reduction rules diverge. Unlike the table of outcomes in \Cref{fig:motivating-example-outcomes}, we can use the formal model to explain \emph{why} a language produces the output it does. The branching evaluation highlights the fact that languages can produce the same output for different reasons. For example, \CSharp and Swift both produce \code|AC|, however, \CSharp does so because the runtime loop terminates running tasks and Swift does so because it cancels unawaited \tasks at the end of their scope. In the evaluation context of \code|ex2|, these semantics produce the same output, but in a different context they would likely lead to different outputs. The branching evaluation also highlights that diverging branches can lead to the same output.

In sum, we presented an illustrative subset of our formal model, constructed as a series of language extensions atop the core \lc (\Cref{fig:langs:lineage}). For each rule---\semrule{Async-App}, \semrule{Spawn}, \semrule{Await}, and the cancellation rules---we marked where the language variants diverge, coloring each fragment by the design dimension it decides. We use the formal model to explain diverging output of real-world examples, illustrated in \Cref{fig:explain-tree} for \code|ex2| from \Cref{fig:motivating-example}. The included artifact provides the full Redex model that we hope can serve as common ground when communicating asynchronous semantics between language communities.

\section{Related Work}\label{sec:relatedwork}

The study of asynchronous programming sits at the intersection of several bodies of literature. We organize related work into three groups: foundational models of concurrency from which asynchrony draws its semantics, the development of deferred-value abstractions that gave asynchrony its first concrete programming primitives, and the functional and monadic tradition that supplied the compositional framework underlying async/await.

The earliest formal treatments of concurrency were not primarily concerned with asynchrony as we have defined it, but they established the semantic vocabulary on which later work depends. The first formalized models were the actor model~\cite{baker_incremental_1977,agha_actors_1986} and communicating sequential processes~\cite{hoare_communicating_1978}. The $\pi$-calculus extended this work by allowing first-class channels that can also be sent between processes~\cite{milner_calculus_1992}.

Deferred value abstractions have been introduced in many forms and under many aliases. \citet{baker_incremental_1977} introduced the term \textit{future,} which was also used by Multilisp, which extended Scheme with several forms for expressing concurrency~\cite{halstead_multilisp_1985}. An independent exploration of demand-driven evaluation by \citet{friedman1976cons} brought about the lazy \textit{cons} operator. \citet{miller_concurrency_2005} defined \textit{promise pipelining:} rather than blocking to receive a value, a program could register a continuation that would execute upon resolution.

Purer functional languages asked not how to represent a deferred value, but rather how to compose effectful computations while maintaining referential transparency. \citet{jones_concurrent_1996} built concurrent Haskell, which was later extended with shared-state concurrency~\cite{harris_composable_nodate}. The monadic model influenced \FSharp, the first major async/await design, where the compiler automatically rewrote a block of code written in direct-style into its continuation-passing equivalent~\cite{syme2011async}.

In addition, \citet{bierman_pause_2012} published a formal model for Featherweight \CSharp (v5) from which we drew much inspiration. The term ``structured concurrency'' emerged from practitioners in early 2016~\cite{sustrik_getting_2016-1,sustrik_getting_2016,sustrik_structured_2025}, but the idea of structured and automatic resource management was introduced in Racket's \emph{custodians}~\cite{flatt:1999custodians}.

\section{Discussion}\label{sec:discussion}

The book is not closed on the design of straight-line asynchrony. Communities are actively exploring extensions to the core async/await semantics presented in this work.


\citet{elizarov_kotlin_2021} use the \emph{suspend} keyword in Kotlin to define async functions. However, in lieu of an explicit await keyword, the language automatically introduces suspension points where necessary. They argue that this approach avoids forgotten awaits, and further that it improves the look-and-feel of calling a suspending function, as it looks identical to calling a non-suspending [synchronous] function. However, this claim, and other consequences of this design, are not accompanied by an evaluation.

The Zig programming language is in the process of stabilizing its ``colorblind'' async/await~\cite{kelley_zigs_2025,cro_what_2020}. The core idea is to parameterize all code over an \abbr{i/o} implementation; libraries can write code to take advantage of asynchrony when available, and library clients can pass any \abbr{i/o} implementation they have available, including a synchronous variant. This approach makes it easier to interoperate sync and async code, but introduces new risks where libraries written explicitly for asynchrony can deadlock and/or panic at runtime if asynchrony is unavailable.

\Cpp{}20 provides a flexible implementation of coroutines that does not ascribe a semantics to its async/await~\cite{isoiec_jtc1_sc22_wg21_technical_2017}. We cannot attribute \Cpp to any particular design point in the taxonomy provided in \Cref{tab:dims-overview} because each axis is configurable. Although elegant and neutral, the choice of full programmability makes each library an async \abbr{dsl}; knowledge transfer between projects within the same language becomes exceedingly difficult.

\paragraph{Where do we go from here?}

We set out to examine the space of straight-line asynchrony, ourselves confused by some statements we read in individual language descriptions. In particular, as we note in \Cref{s:sync-quotes-from-langs}, these languages motivate async from a perspective of similarity to synchrony, which we found difficult to reconcile. Our design space exploration demonstrates the numerous ways in which this analogy fails.

The principal use for our exploration is to help developers. Developers transitioning from one language to another carry their prior knowledge to make sense of the new in terms of the old. It is then especially confusing when the new language has the same keywords and similar description as the old one, but with a different semantics. We believe that our articulation of the design space, such as \Cref{tab:dims-overview}, can help both learners and their educators talk more effectively about how to translate knowledge between languages.
We also believe that our exploration can help language designers in building new features for straight-line asynchrony. Our design space can help enumerate key design dimensions, preventing important details from slipping through the cracks of the design process. Our design space can also help designers more effectively compare and contrast their design against related languages.


An open challenge for the future is to build a body of empirical data that can support designers in reasoning about the trade-offs of different approaches. For example, how much overhead does \seager async function application incur versus \eager, and how much more concurrency does it gain? To what extent does \simultaneous cancellation prevent realistic bugs that would occur with a \bup cancellation strategy? Which of these semantics best matches developer expectations for the behavior of asynchronous code? Which decisions lead to more errors? We hope that this design space exploration can inspire practitioners and academics alike to build a better understanding of the evolving world of asynchronous programming.

\newpage

\section*{Data-Availability Statement}

The software artifact for this paper is publicly available~\cite{zenodo:artifact} and contains three main components:
\begin{enumerate}

  \item All the code associated with the examples in the paper,
    runnable in their respective languages.

  \item The full suite of Redex models providing the described
    semantics.

  \item A differential fuzzer to test each model against its real-world counterpart.

  \end{enumerate}

\begin{acks}
  We want to thank the many people who provided feedback on this work: Luke Wagner and Alex Crichton for detailing the WASM component model, John McCall for helping us understand the nuances of Swift's semantics, Akshay Narayan and Malte Schwarzkopf for their close readings of the text and thoughtful feedback that helped scope and frame the paper, and the members of the Brown PLT and Cognitive Engineering Lab groups for testing artifacts and spontaneously answering the question \emph{What do you think this program prints?}.

  This work is partially supported by US NSF grant CCF-2227863 and by the DARPA under Agreement No.\@ HR00112420354. Any opinions, findings, and conclusions or recommendations expressed in this material are those of the authors and do not reflect the views of our funders.
\end{acks}

\bibliography{bibs/bibfile-mod,bibs/aux}


%


\end{document}